\documentclass[printer]{aa}

\usepackage{graphicx}
\usepackage{txfonts}

\usepackage{amssymb}
\usepackage{mathtools}
\usepackage{lscape}
\usepackage{natbib}
\usepackage{gensymb}
\usepackage{adjustbox}
\usepackage{multirow}
\usepackage{epstopdf}
\usepackage{tabulary}
\usepackage{tabularx}
\usepackage{hhline}
\usepackage{natbib}
\bibpunct{(}{)}{;}{a}{}{,} 
\usepackage[T1]{fontenc}

\begin{document}

   \title{Submillimetre-sized dust aggregate collision and growth properties}

   \subtitle{Experimental study of a multi-particle system on a suborbital rocket}

   \author{J. Brisset
          \inst{1} \inst{2} \inst{3}
          \and
          D. Hei\ss elmann \inst{1}
	\and
	S. Kothe \inst{1}
	\and
	R. Weidling \inst{1}
	\and
	J. Blum \inst{1}
          }

   \institute{Institut f{\"{u}}r Geophysik und extraterrestrische Physik, Technische Universit\"at Braunschweig, Mendelssohnstr. 3, 38106 Braunschweig, Germany\\
              \email{j.brisset@tu-bs.de}
         \and
             Max Planck Institute for Solar System Research, Justus-von-Liebig-Weg 3, 37077 G{\"o}ttingen, Germany
         \and
	    Present address: Center of Microgravity Research, University of Central Florida, 4111 Libra Drive, Orlando FL-32816, USA
             }


 
  \abstract
   {In the very first steps of the formation of a new planetary system, dust agglomerates grow inside the protoplanetary disk that rotates around the newly formed star. In this disk, collisions between the dust particles, induced by interactions with the surrounding gas, lead to sticking. Aggregates start growing until their sizes and relative velocities are high enough for collisions to result in bouncing or fragmentation. With the aim of investigating the transitions between sticking and bouncing regimes for colliding dust aggregates and the formation of clusters from multiple aggregates, the Suborbital Particle and Aggregation Experiment (SPACE) was flown on the REXUS 12 suborbital rocket.}
   {The collisional and sticking properties of sub-mm-sized aggregates composed of protoplanetary dust analogue material are measured, including the statistical threshold velocity between sticking and bouncing, their surface energy and tensile strength within aggregate clusters.}
   {We performed an experiment on the REXUS 12 suborbital rocket. The protoplanetary dust analogue materials were micrometre-sized monodisperse and polydisperse SiO$_2$ particles prepared into aggregates with sizes around 120 $\mu$m and 330 $\mu$m, respectively and volume filling factors around 0.37. During the experimental run of 150~s under reduced gravity conditions, the sticking of aggregates and the formation and fragmentation of clusters of up to a few millimetres in size was observed.}
   {The sticking probability of the sub-mm-sized dust aggregates could be derived for velocities decreasing from $\sim$22 to 3~cm~s$^{-1}$. The transition from bouncing to sticking collisions happened at 12.7$_{-1.4}^{+2.1}$~cm~s$^{-1}$ for the smaller aggregates composed of monodisperse particles and at 11.5$_{-1.3}^{+1.9}$ and 11.7$_{-1.3}^{+1.9}$~cm~s$^{-1}$ for the larger aggregates composed of mono- and polydisperse dust particles, respectively. Using the pull-off force of sub-mm-sized dust aggregates from the clusters, the surface energy of the aggregates composed of monodisperse dust was derived to be 1.6$\times10^{-5}$~J~m$^{-2}$, which can be scaled down to 1.7$\times10^{-2}$~J~m$^{-2}$ for the micrometre-sized monomer particles and is in good agreement with previous measurements for silica particles. The tensile strengths of these aggregates within the clusters were derived to be 1.9$_{-1.2}^{+2.2}$~Pa and 1.6$_{-0.6}^{+0.7}$~Pa for the small and large dust aggregates, respectively. These values are in good agreement with recent tensile strength measurements for $\sim$mm-sized silica aggregates.}
   {Using our data on the sticking-bouncing threshold, estimates of the maximum aggregate size can be given. For a minimum mass solar nebula model, aggregates can reach sizes of $\sim$1~cm.}


\maketitle

\section{Introduction}
\label{s:intro}

The formation of planets around a young star takes place in the remains of accreting clouds of gas and dust, which form disks around their pre-main-sequence stars \citep{weidenschilling_cuzzi1993PP3, weidenschilling2000Icarus, dominik_et_al2007PPV}. In these so-called protoplanetary disks (PPDs), the dust has condensed into solid grains that range in size from sub-$\mu$m- to $\mu$m \citep{Bouwman_et_al2007ApJ}. Several stochastic (Brownian motion) and aerodynamic processes (e.g. gas drag and turbulence) influence the dynamics of these grains in the nebula and induce velocity differences between them, which lead to collisions and initiate the growth of aggregates \citep[e.g. see the reviews by][]{johansen_et_al2014PP,testi_et_al2014PP,blum_wurm2008ARAA}.

For grain sizes of about one micrometre in these PPDs, collisions always lead to grain growth through van der Waals forces \citep{blum_wurm2008ARAA}. This is the so-called "hit-and-stick" behaviour. The growing dust aggregates, however, soon become big enough to leave this regime. As their sizes increase, their relative velocities increase accordingly \citep{weidenschilling1977MNRAS} and collisions then lead to restructuring, bouncing, or even fragmentation, rendering processes that allow further grain growth more complex \citep{blum_wurm2008ARAA,guettler_et_al2010A&A,zsom_et_al2010AA}. Reproducing the complete growth from $\mu$m-sized dust grains to km-sized planetesimals is an ongoing endeavour. Amongst the many processes to account for in PPDs, e.g. concentration of dust aggregates in streaming instabilities \citep{youdin_goodman2005} or gravitational instabilities of dense dust layers \citep{johansen_et_al2009ApJ,johansen_et_al2012AA}, the collision behaviour of growing dust aggregates also serves as a crucial input to PPD models \citep{windmark_et_al2012AAa,windmark_et_al2012AAb,garaud_et_al2013ApJ}. Knowing under which conditions and by which processes dust growth is possible is of utmost importance for understanding and simulating the first stages of the formation of planetary bodies.

While the exact conditions present in PPDs are still under discussion, investigating the collision behaviour of dust aggregates, without any influence of gas, temperature nor other disturbances is already very useful for developing dust evolution models. In order to further our understanding of the processes underlying the dust collision behaviour, many experiments as well as numerical simulations have been (and still are) performed \citep[see reviews by][]{dominik_et_al2007PPV,  blum_wurm2008ARAA}. \citet{guettler_et_al2010A&A} compiled the results of several laboratory and microgravity dust collision experiments into a model predicting the outcome of a collision between two arbitrary dust aggregates, depending on their mass, porosity, and relative velocity. In particular, the transition between sticking and bouncing collisions is an area of interest because the number of experimental investigations for macroscopic dust aggregates probing into this transitions is very small. Indeed, depending on the collision outcomes, simulations of dust aggregate growth in PPDs can have very different results, even leading to stalling growth at centimetre sizes \citep{wada_et_al2011ApJ,zsom_et_al2010AA}. As far as we know, only two dust-aggregate collision experiments were conducted with particle sizes and at relative velocities adequate to observe this sticking-bouncing transition. \citet{weidling_et_al2012Icarus} observed mm-sized silicate dust aggregates colliding at velocities of $\sim$1~cm~s$^{-1}$ and \citet{kothe_et_al2013Icarus} observed 100~$\mu$m-sized aggregates at $\sim$10~cm~s$^{-1}$ \citep[see][for an update of the collision model]{kothe_et_al2013Icarus}. These experiments were performed with many-particle systems, which also display "group" effects like clustering \citep{kothe_et_al2013Icarus}.

In this context, the experimental investigation presented in this paper was designed to gather additional data on collisions between sub-mm-sized dust aggregates. One of the challenges to observing dust collisions in many-particle systems in a range of velocities down to 10~cm~s$^{-1}$ and below is the necessity to conduct these experiments under reduced gravity conditions. The flight profile of suborbital rockets allows for an experiment to remain in such reduced gravity conditions (down to 10$^{-3}g$) for several minutes. This paper presents the Suborbital Particle Aggregation and Collision Experiment (SPACE) that flew on the REXUS 12 rocket (Rocket EXperiments for University Students, a joint project between DLR -- Deutsches Zentrum f{\"u}r Luft-und Raumfahrt -- and SNSB -- Swedish National Space Board) in March 2012 and the analysis of the gathered data. In Section~\ref{s:setup} we present the experimental setup and flight performance as well as the dust aggregates we chose to observe. Section~\ref{s:results} presents the results of the analysis of the gathered data. In Section~\ref{s:discussion} we discuss these results and their applicability to dust collisions in protoplanetary disks.

\section{Experimental setup}
\label{s:setup}
This section presents the experiment hardware setup and the scientific data gathered by the SPACE experiment, which flew onboard the REXUS 12 suborbital rocket in March 2012. More details about the experimental hardware and rocket can be found in \citet{brisset_et_al2013RSI}.

\subsection{Experiment hardware}
\begin{figure}[t]
  \begin{center}
  \includegraphics[width = 0.45\textwidth]{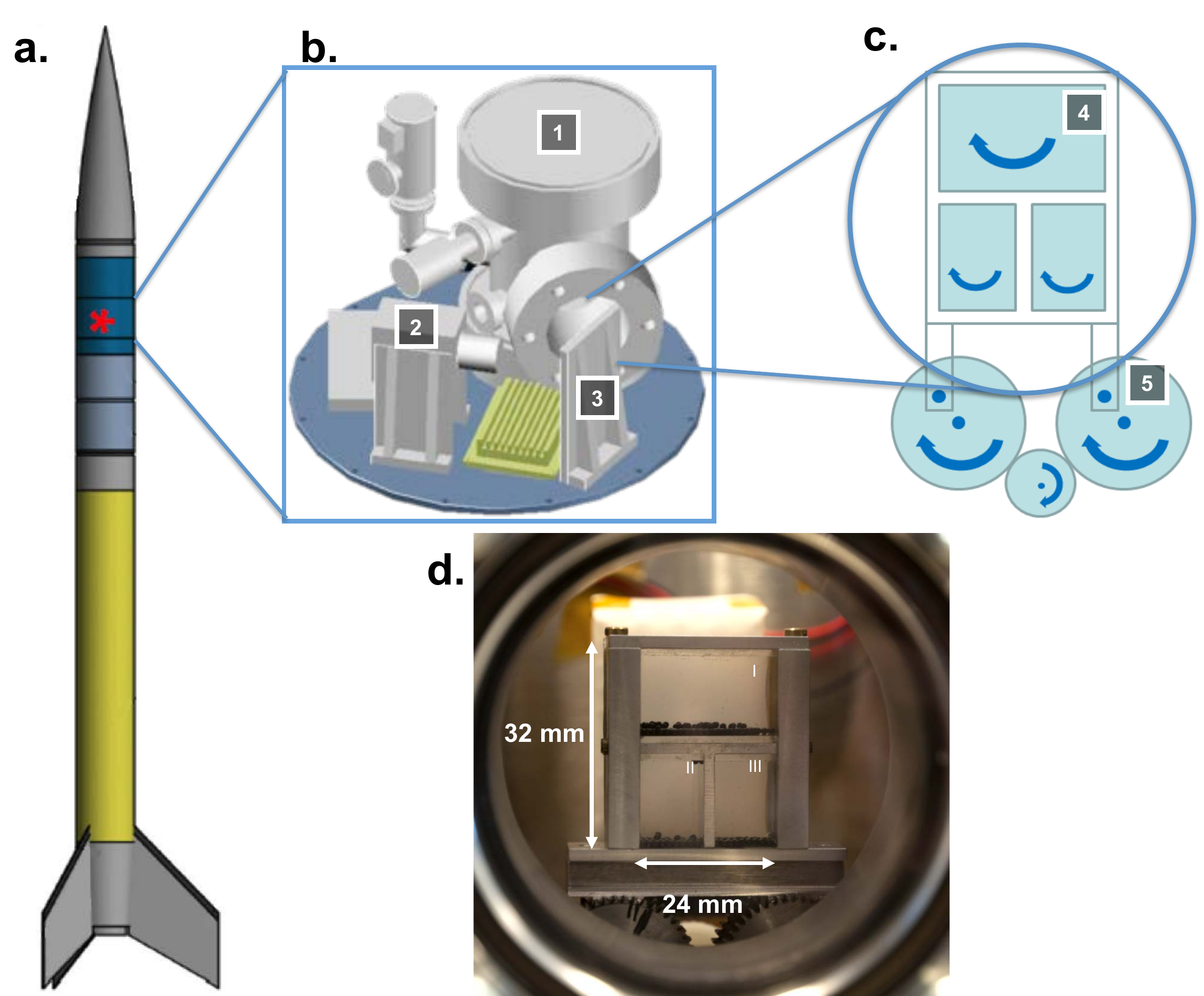}
 \caption{Experiment cells inside the rocket payload: a. Schematic of the REXUS 12 rocket with the position of the SPACE module marked by a red asterisk. b. CAD drawing of the SPACE payload including the vacuum chamber (1) and the camera (2) observing the interior of the chamber through a mirror (3). c. Schematic view of the glass experiment cells (5) installed on the shaking mechanism (6). The gears of the shaking mechanism activated by a motor induced the circular motion of the cells indicated on the figure by blue arrows. The circle indicates the limits of the vacuum chamber viewport. d. Photograph of the SPACE experiment cells inside the vacuum chamber and seen through the view port. The types of dust inserted into each cell were $\sim$120~$\mu$m-sized monodisperse in the larger cell (I) and $\sim$330~$\mu$m-sized poly- and mono-disperse in each of the smaller cells (II and III, resp.).}
 \label{f:cells}
 \end{center}
\end{figure}

The purpose of the experiment being to observe dust-aggregate collisions in a multi-particle system, we placed sub-mm-sized dust aggregates into an evacuated glass container and recorded their behaviour during the extended microgravity phase with a high-speed camera. The particle container was divided in three cells, two smaller ones of dimensions 11$\times$10$\times$15 mm$^3$ and a bigger cell with dimensions of 24$\times$10$\times$15 mm$^3$ (see Figure \ref{f:cells}). These cells allowed for the observation of different types of dust aggregates during a single rocket flight. The inner glass walls of these cells were coated with an anti-adhesive nano-layer to reduce their sticking efficiency with the dust aggregates \citep[see Figure 5 in][for more details]{brisset_et_al2013RSI}.

In these types of microgravity dust-aggregate collision experiments, experience has shown that in order to obtain and maintain a uniform spatial distribution of the aggregates in the cell volume (e.g. against microgravity disturbances by air drag acting on the rocket), a shaking of some kind has to take place. Indeed, if a many-particle system is left on its own in microgravity, it kinetically "cools" down quite rapidly, losing energy with each collision \citep{haff1983JFM} and leading to a lower collision frequency among the aggregates. Shaking also allows for the deagglomeration of clusters formed during the experiment preparation as well as for a control over the mean collision velocity between the aggregates. As suborbital rockets are subject to residual accelerations (see Appendix~\ref{a:res_acc}), shaking of the experiment was an essential part of the hardware setup. The two main microgravity disturbances during flight were the atmospheric drag on the rocket body, which acts in the direction of flight, and the spinning centrifugal force, which acts in the outward radial direction and perpendicular to the direction of flight. Therefore, we decided to shake the dust cells in a circular motion along these two directions (see Figure~\ref{f:cells}). The radius of the cell rotation being 1~mm, shaking frequencies between 5 and 30~Hz led to linear wall velocities between about 3 and 22~cm~s$^{-1}$.

\begin{figure}[t]
  \begin{center}
  \includegraphics[width = 0.45\textwidth]{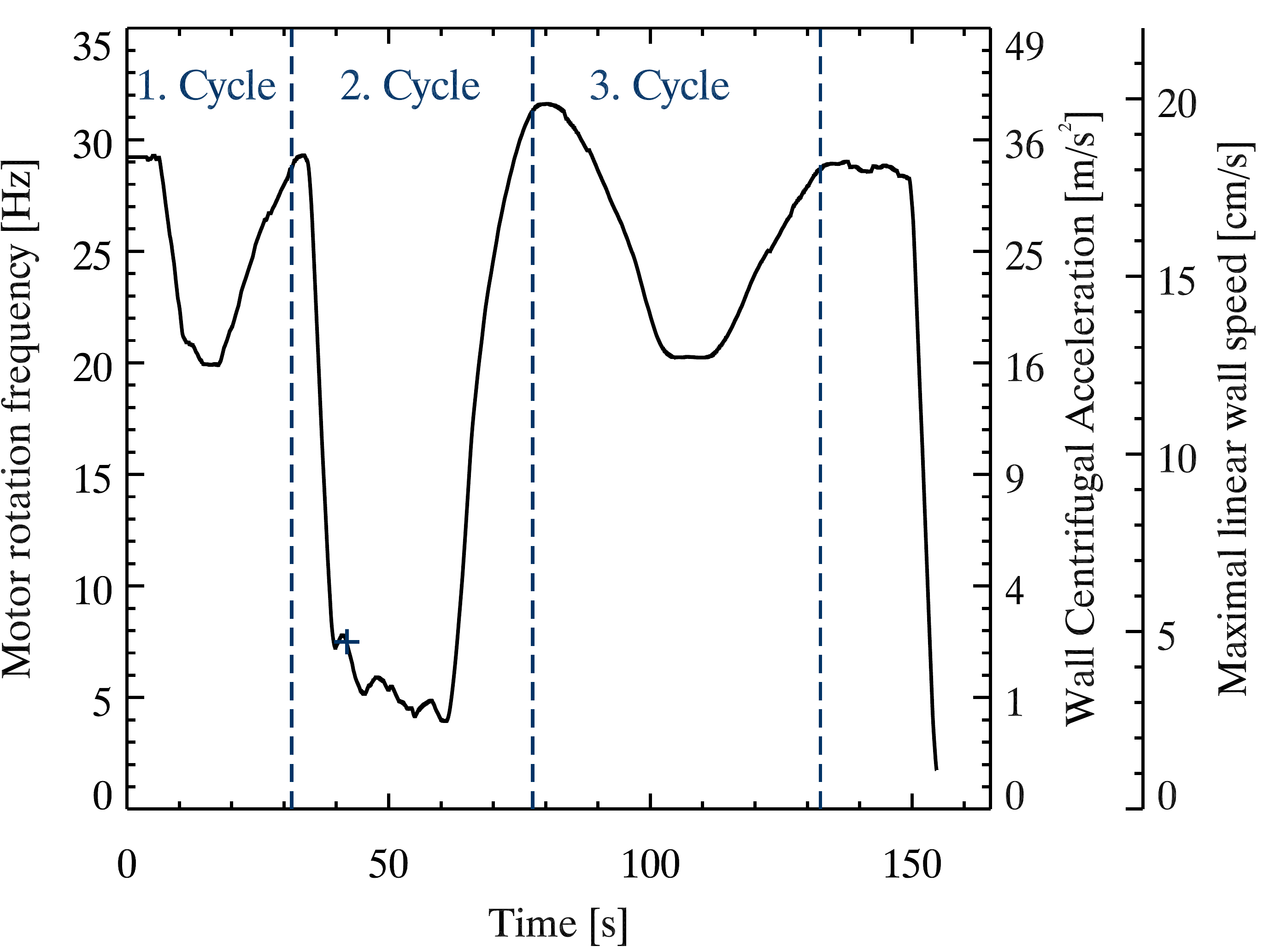}
 \caption{Measured rotation frequency of the particle cells during the experimental run of SPACE onboard the REXUS 12 rocket. Time is measured from the first recorded frame ($\sim$ 100 s after lift-off of the rocket). The corresponding wall centrifugal acceleration and maximal linear speed are indicated. The apogee of the rocket trajectory is indicated by a "+" sign.}
 \label{f:motor_profile}
 \end{center}
\end{figure}

The shaking-velocity profile included three cycles, each composed of a fast shaking period of 5 to 10~s and a longer slow shaking period of 15 to 25~s with transition ramps between them (see Figure~\ref{f:motor_profile}). This shaking profile was designed to observe three agglomeration phases (during ramping down the rotation speed of the cells) and three de-agglomeration phases (during ramping up the rotation speed of the cells). The slowest and longest shaking phase was scheduled around the apogee of the rocket trajectory, when the quality of the microgravity was at its best (i.e. the residual accelerations were lowest). As shown in Figure~\ref{f:motor_profile}, the shaking of the experiment kept the particle systems in a relatively high acceleration environment. However, this profile choice allowed for the observation of the dust aggregates over a continuous range of speeds, both decreasing and increasing. In addition, the slow shaking speeds of cycle 2 kept the particle systems under 1$g$ for a significant amount of time ($\sim$30~s) during which the transition from a bouncing to a sticking regime were observed.

To be able to observe the aggregate collisions without the influence of gas, the glass containers were placed in a vacuum chamber, which was evacuated during the experiment. The pressure was logged during the rocket flight and remained below 10$^{-4}$~mbar (10$^{-2}$~Pa) for the duration of the experiment.

The evolution of the dust-aggregate ensembles in the three glass cells were simultaneously and continuously recorded with an autonomous high-speed camera with a chip size of 640 $\times$ 480 pixels, resulting in a spatial resolition of 56.6~$\mu$m/px, at a recording speed of 170 frames per second.

\subsection{Dust aggregate properties}
\label{s:dust}
In this paper, we will refer to the single dust aggregates (a few 100 $\mu$m in size) introduced into the experiment glass cells before the beginning of the experiment as aggregates, dust aggregates, or monomer aggregates. These aggregates themselves consist of smaller dust grains of $\sim$1 $\mu$m in size, which we will call monomer particles. In their storage container, these monomer particles form aggregates, which we then sieved to the desired size distributions. When several of these dust aggregates stick together during the experiment to form a bigger agglomerate, we will refer to it as a cluster.

The dust analogue material we used in the SPACE experiment was SiO$_2$, identical to that used by \citet{weidling_et_al2012Icarus} and \citet{kothe_et_al2013Icarus}, and frequently used in dust collision experiments \citep[see][and references therein]{blum_wurm2008ARAA}. The material density of SiO$_2$ is $\rho$  =  2000 kg~m$^{-3}$ for the (slightly nano-porous) spherical monomer particles (monodisperse dust) and $\rho$  =  2600 kg~m$^{-3}$ for the irregular monomer particles (polydisperse dust). Further properties can be retrieved from \citet{blum_et_al2006ApJ}. \citet{kothe_et_al2013Icarus} determined the average volume filling factor of the sieved clusters to be $\phi$ =  0.37$^{+0.06}_{-0.05}$, in agreement with the measurements of \citet{weidling_et_al2012Icarus}. Building on \citet{guettler_et_al2009ApJ} and \citet{zsom_et_al2010AA}, we consider this value of the aggregate filling factor to be representative of protoplanetary dust aggregates at the sizes of aggregates studied here. It should be mentioned that a filling factor of $\phi$=0.4 as found by \citet{zsom_et_al2010AA} is representative of aggregates consisting of $\sim$1~$\mu$m silicate particles. However, \citet{kataoka_et_al2013AA} found that 0.1~$\mu$m ice particles form aggregates that are fluffier. \citet{kothe_et_al2013Icarus} also investigated the inner structure of the aggregates with the result that the standard preparation process of the aggregates does not induce any kind of compacted shell, meaning that a homogeneous structure of the aggregates can be assumed \citep[see Figure~4 in][]{kothe_et_al2013Icarus}.

\begin{figure}[tp]
  \begin{center}
  \includegraphics[width = 0.48\textwidth]{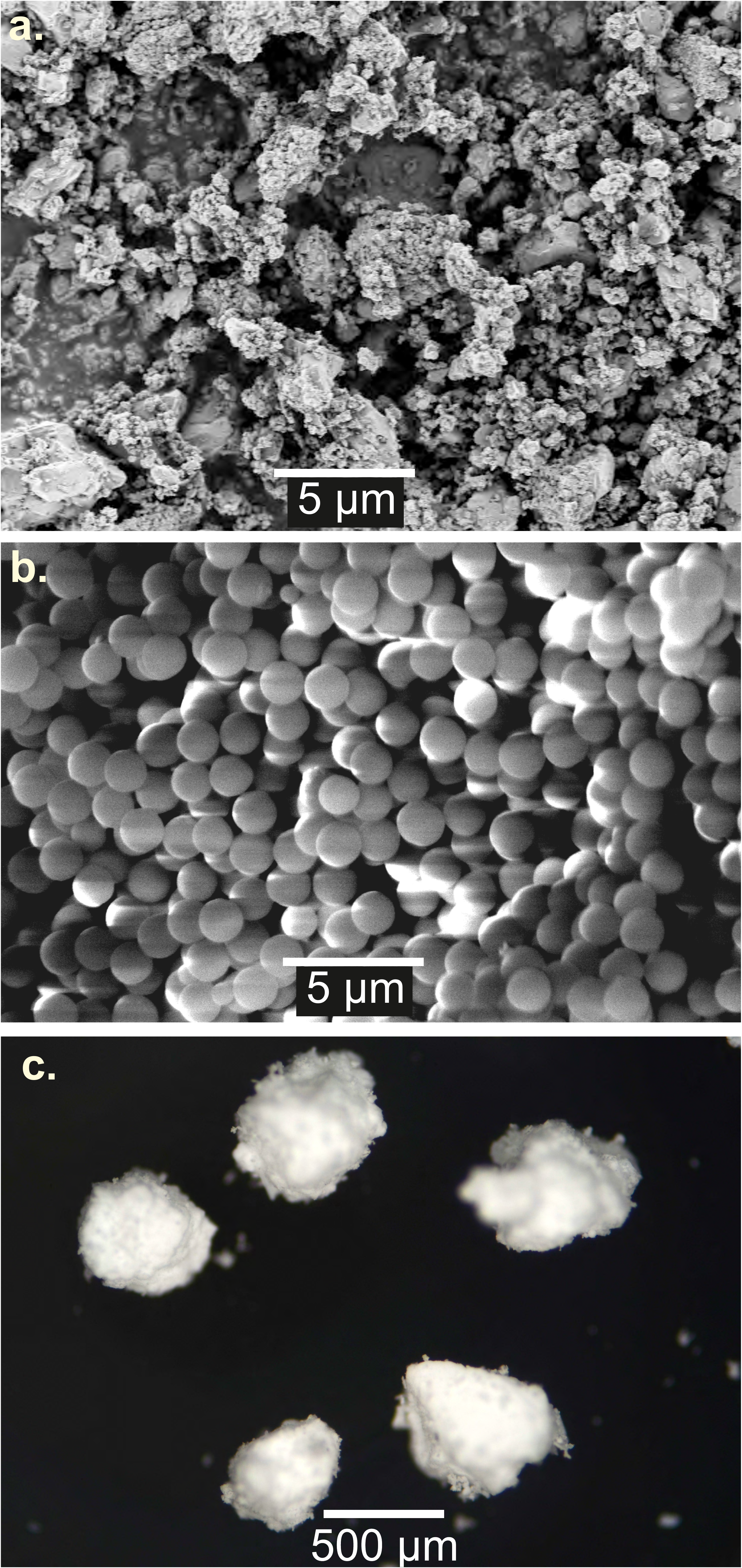}
 \caption{SiO$_2$ dust used in the SPACE experiment: a. SEM image of the polydisperse SiO$_2$ monomer particles (image credit: E. Beitz). b. SEM image of monodisperse SiO$_2$ monomer particles (image credit: E. Beitz). c. Microscope image of aggregates composed of irregular polydisperse SiO$_2$, sieved between 250 and 500 $\mu$m.}
 \label{f:dust_pics}
 \end{center}
\end{figure}

\begin{figure}[tp]
  \begin{center}
  \includegraphics[width = 0.5\textwidth]{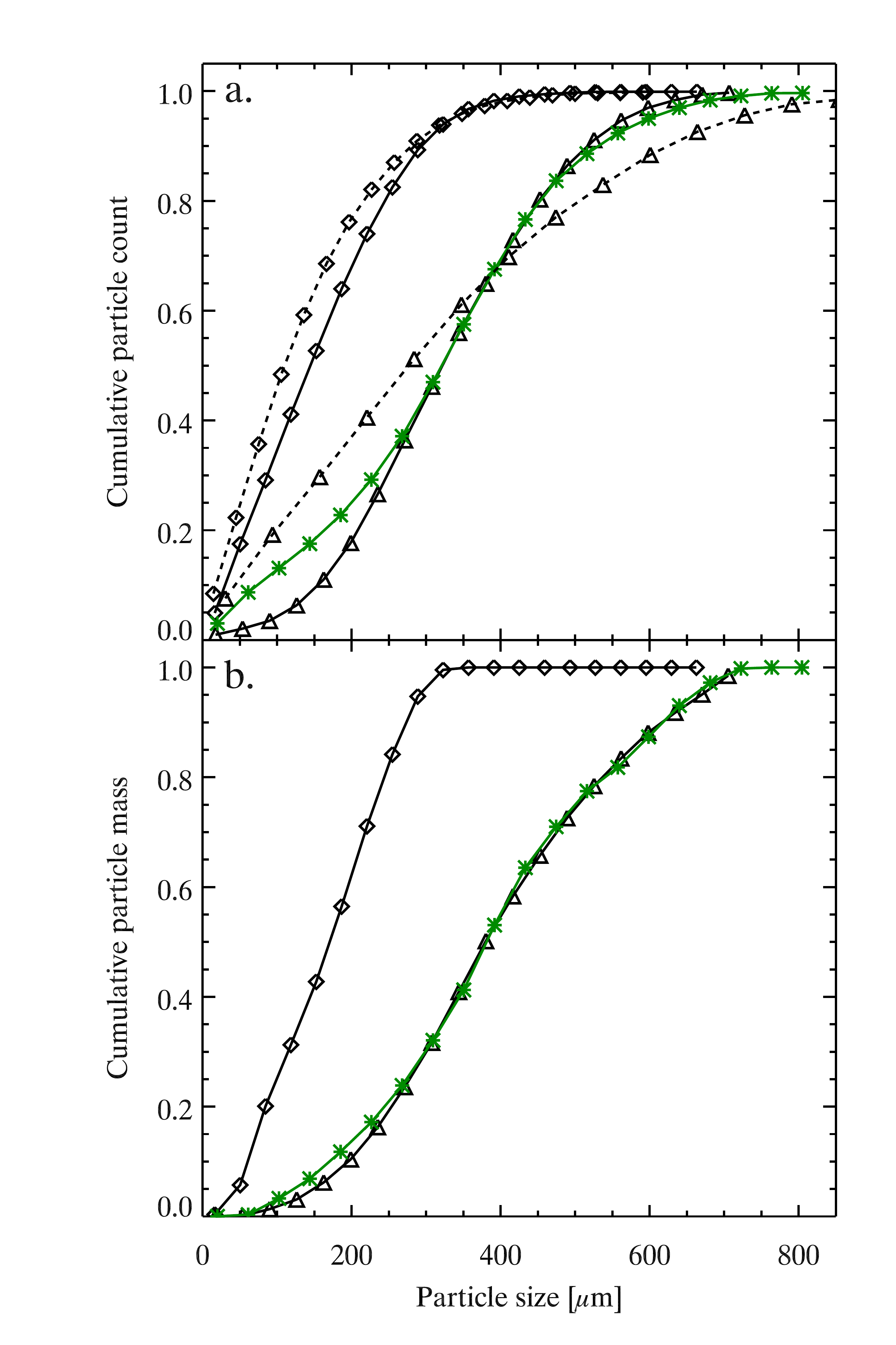}
 \caption{Size and mass distributions of the aggregate samples used in this study. a. Normalized cumulative number of aggregates with a size smaller than the size indicated on the x-axis. b. Normalized cumulative mass of aggregates smaller than the size indicated on the x-axis for the three different SiO$_2$ aggregates used in the SPACE experiment. Black diamonds denote aggregates sieved between 100 and 250 $\mu$m and composed of monodisperse dust particles. Black triangles denote aggregates sieved between 250 and 500 $\mu$m and composed of monodisperse dust particles. Green asterisks denote aggregates sieved between 250 and 500 $\mu$m and composed of polydisperse dust. Dashed lines denote aggregate sizes measured during the experiment run on the rocket.}
 \label{f:dust_distrib}
 \end{center}
\end{figure}

Before the REXUS flight, the two SiO$_2$ samples (monodisperse, see Figure~\ref{f:dust_pics}b., and polydisperse, see Figure~\ref{f:dust_pics}a.) were sieved into two different size distributions, one between 100 and 250 $\mu$m and the other one between 250 and 500 $\mu$m. The mean aggregate size was then measured to be 120~$\mu$m and 330~$\mu$m, respectively (see Figure~\ref{f:dust_distrib}). The dust samples were introduced into the experiment cells as follows (see Figure~\ref{f:cells}b.): 19.2~mg of the spherical monodisperse SiO$_2$ dust and aggregate sizes between 100 and 250 $\mu$m were placed into the bigger glass cell, 15.3~mg of the spherical monodisperse SiO$_2$ dust and aggregate sizes between 250 and 500 $\mu$m were placed into the right (seen from the camera location) smaller cell, and 19.8~mg of the irregular polydisperse SiO$_2$ dust and aggregate sizes between 250 and 500 $\mu$m were placed into the left (seen from the camera location) smaller cell. These quantities are the total masses of dust inserted into each experiment cell.

\subsection{Scientific data} 

\begin{figure}[t]
  \begin{center}
  \includegraphics[width = 0.48\textwidth]{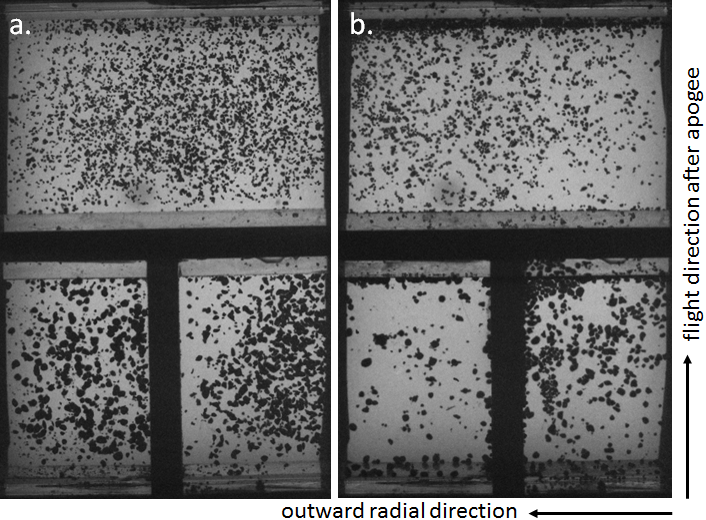}
 \caption{Two image frames recorded during the SPACE experiment. a. At the very beginning of the experimental run, when all the dust aggregates were free-flying. b. During a slow shaking phase of the experiment (see Figure~\ref{f:motor_profile}). In slow shaking phases, the preferential accumulation of aggregates in the cells' left upper corner due to microgravity disturbances (see \ref{a:res_acc} for details) is visible. The flight direction of the rocket (i.e. direction of gas drag) and the outward radial direction (i.e. direction of centrifugal force) are indicated.}
 \label{f:frames}
 \end{center}
\end{figure}

The SPACE experiment performed nominally during the RE-XUS 12 flight. Two example frames of the gathered scientific data can be seen in Figure \ref{f:frames}. During the experiment run, the observed collisions between dust aggregates formed clusters on the inner walls of the glass containers. We show in Appendix~\ref{a:mc} that this was due to the non-negligible residual acceleration increasing the number of collisions between the dust aggregates and the container walls. In the two directions in which the experiment cells were shaken, this had no major influence on the dust-glass interactions, as the shaking accelerations were almost 3 orders of magnitude stronger than the flight residual accelerations (see Appendix~\ref{a:res_acc}). However, no shaking was applied in the line of sight of the camera. The increased number of dust-glass collisions due to the residual flight accelerations led to the sticking of a small number of aggregates to the cell walls with the very low sticking probability of about 0.5 \% (the cell walls were coated anti-adhesively, see Appendix~\ref{a:mc}). These aggregates sticking to the walls acted as seeds for further cluster growth (see Appendix~\ref{a:analysis}). 

During the first and third shaking phases (see Figure \ref{f:motor_profile}), the aggregate speeds were too high (>10~cm~s$^{-1}$) to detect a significant change in collision behaviour. In the second phase, however, a clear transition from the bouncing to the sticking regime could be observed while the shaking speed was being ramped down (<5~cm~s$^{-1}$). During the speed ramp-up of this same phase, the dismantling of the formed clusters could be observed.

\section{Results}
\label{s:results}

\subsection{Growing clusters on the cell walls}
This section describes how cluster formation on the glass walls of the experiment cell can deliver information on dust aggregate properties and growth. To support the fact that clusters grew through aggregate-cluster collisions instead of aggregate-glass collisions we performed a series of Monte Carlo simulations in which we randomly deposited particles on a surface, representing the cell glass walls (see Appendix~\ref{a:mc}). As detailed in Appendix~\ref{a:mc}, we performed numerous runs with varying the number of aggregates as well as their sticking probability to the glass walls. In each case, the sticking probability in an aggregate-cluster collision was set to unity. We then compared the number and morphology of the forming clusters to those found in the experiments. 

The result of this investigation was that a good match between simulation and experiment was found when the probability of a dust aggregate to stick to the glass upon collision was as low as 0.5~\%, which shows that the applied anti-adhesive coating was intact over 99.5~\% of the surface but showed defects at places where the impinging aggregates could stick. This means that it was not an enhanced glass sticking efficiency that led to clusters on the cell walls but indeed the enhanced number of collisions of the aggregates on the glass, induced by residual accelerations during the rocket flight (see Appendix~\ref{a:res_acc}). It can be concluded that an individual dust cluster was not growing on the glass directly but on a single dust aggregates that stuck to the glass and served as a seed for further impinging aggregates. Accordingly, clusters in the SPACE experiment grew by collisions between dust aggregates and the clusters on the wall so that the further data analysis will deliver information on the collision and sticking properties of these aggregates.

The difference between collisions among free-flying aggregates and those between aggregates and clusters sticking to a glass wall is that the cluster mass in the latter case can be considered as infinite compared to a finite mass of free-flying clusters. Aggregate collisions simulated by \citet{wada_et_al2011ApJ} show for example that a low-velocity collision ($<$2~m~s$^{-1}$ for their icy aggregates) only locally affects the involved aggregates (see their Figure~5) so that their surface properties are unchanged elsewhere. A collision of a free-flying aggregate with a cluster or an aggregate on the wall is therefore equivalent to a collision with a free-flying cluster or aggregate, except for a slight difference (up to a factor 2 in the extreme case of two equal-size aggregates colliding) in the reduced mass, which is a negligible effect for clusters consisting of dozens of aggregates or more.

\subsection{Determination of the aggregate collision velocities}
\label{s:velocity}
During most phases of the experiment run (fast cell shaking), the optical depth in the SPACE suborbital flight experiment was too high and the image-recording frame rate too low to track aggregates individually. In these phases, it was not possible to determine collision velocities by following the aggregate positions over time. When decreasing the shaking speed under the sticking-bouncing transition, the experiment cells almost instantly got depleted in free-flying aggregates, which were then not available for individual aggregate tracking. Hence, only a very narrow span of time ($\sim$5~s during the shaking speed ramp-down of phase 2) allowed for aggregate tracking.

We therefore adopted a statistical approach to derive the aggregate velocities according to the shaking frequency of the experiment cells. The mean velocities derived from this statistical analysis for free-flying aggregates were then compared to a discrete set of aggregate velocities directly measured from the experiment data, when aggregate tracking was possible (shaking ramp-down of phase 2). 

For this statistical approach, the residual accelerations during flight (see Appendix~\ref{a:res_acc}) were neglected and the velocity of the dust aggregates was assumed to be solely induced by collisions with the container walls. The mean free path of the aggregates inside their cell volume was calculated by $\lambda$~=~1/($n~\sigma$) with $n$ being the number density of particles in the experiment cell and $\sigma$ their mean mutual collision cross section, respectively. For the smaller size distribution counting $\sim$4240 aggregates of a mean diametre of 120~$\mu$m and a cell volume of 15$\times$10$\times$24~mm$^3$, the mean free path was calculated to be $\lambda$~=~16~mm. For the larger size distributions counting $\sim$375 aggregates of a mean diametre of 320~$\mu$m and a cell volume of 15$\times$10$\times$11~mm$^3$, the mean free path was calculated to be 13~mm. Thus, in both cases the mean free path of the aggregates was comparable with the size of the cell. Considering collisions between aggregates and the cell walls to be perfectly inelastic, and knowing that the residual gas pressure in the cells was below 10$^{-4}$~mbar during the experiment run (i.e. the influence of gas drag on the aggregate trajectories can be neglected), the free-flying aggregates were assumed to obtain and retain the maximum linear wall velocity until they collided with another aggregate or cluster on the cell walls.

\begin{figure}[t]
  \begin{center}
  \includegraphics[width=0.48\textwidth]{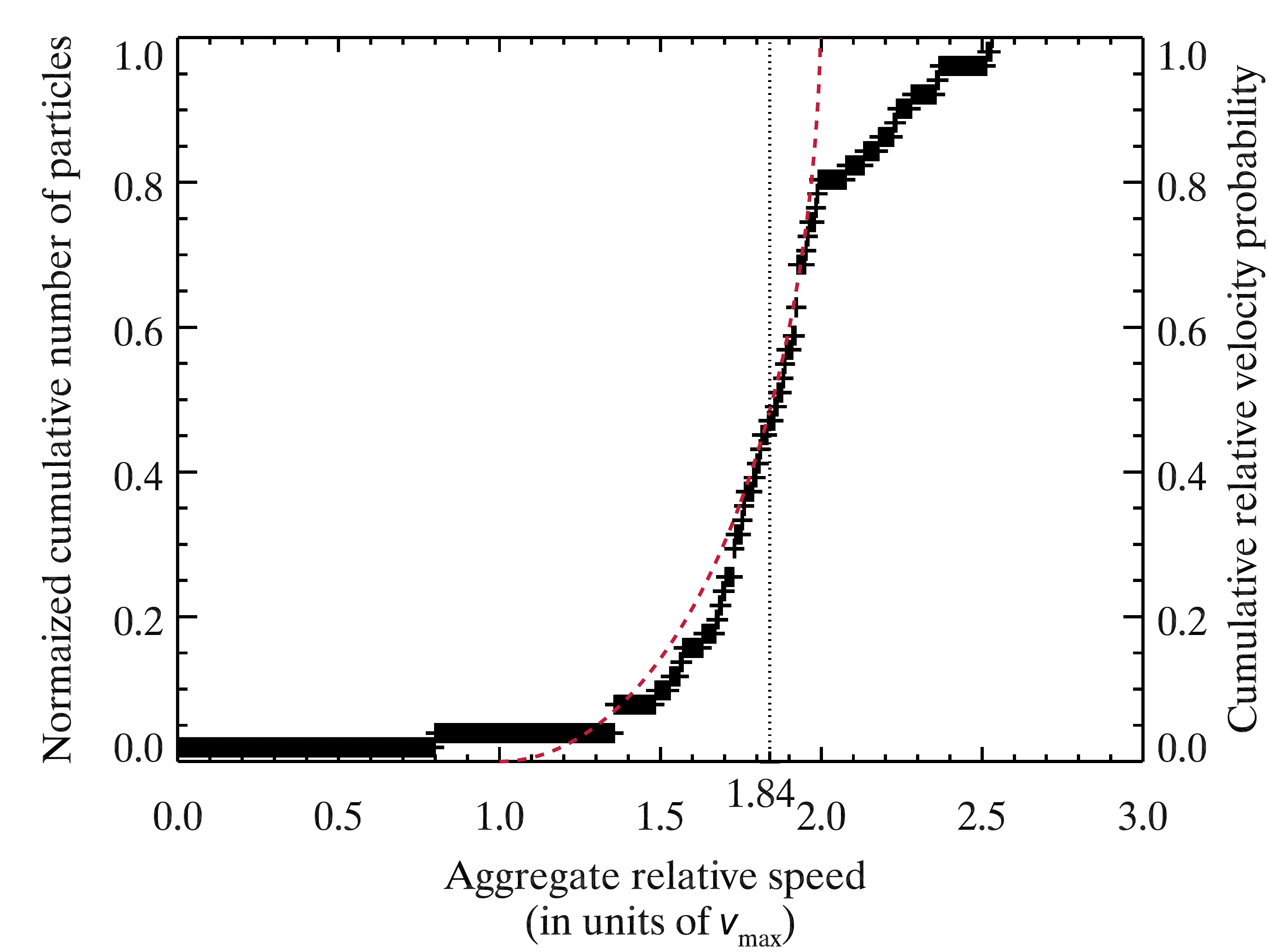}
 \caption{Resulting relative collision velocity probability distribution function between an in-flying aggregate at $v_{\textrm{max}}$ and a rotating cluster on the wall with linear velocity $v_{\textrm{max}}$ (red dashed line). The cumulative distribution of 51 relative aggregate velocities measured during the shaking speed ramp-down of phase 2 is shown ("+" signs). The Gaussian mean (location of the Gaussian distribution peak) and standard deviations of the measured relative velocities are at 1.84$_{-0.21}^{+0.30}v_{\textrm{max}}$ (vertical dashed line).}
 \label{f:convoled}
 \end{center}
\end{figure}

To estimate the mean aggregate speed in the experiment cells during fast shaking phases, we considered that the configuration of the SPACE experiment, with clusters rotating at the speed of the cell walls (see Figure~\ref{f:motor_profile} for the shaking profile), was similar to the one studied by \citet{weidling_et_al2009ApJ} where an aggregate of known velocity collides with a wall moving vertically and sinusoidally with time. Therefore, we applied their method to statistically determine the probability of a certain relative velocity between the incoming free-flying aggregate and the rotating wall cluster. This method relies on the increased probability of aggregates to collide with clusters that are coming towards them in their current circular motion (clusters are following the motion of the experiment cells, see \cite{brisset2014_diss} for details). The resulting probability distribution function can be seen in Figure~\ref{f:convoled} (cumulative, red dashed line). The most frequent collision velocity was at twice the maximum linear wall velocity, 2$v_{\textrm{max}}$. 

To confirm this modelled probability distribution, we measured the velocities of 51 free-flying aggregates by direct tracking during the short time interval in which this was possible (shaking speed ramp-down of phase 2). The obtained relative velocity distribution for collisions with clusters on the cell walls is shown in Figure \ref{f:convoled} ("+" signs). About 15\% of the relative speeds were higher than 2$v_{\textrm{max}}$. This can be attributed either to semi-elastic collisions with the cell walls, or to extra kinetic energy stored in aggregate rotation that was released during the aggregate-wall collision. For the rest of the measured aggregate speeds, the velocity distribution is in very good agreement with the simple model. We can therefore assume a shaking speed dependant velocity distribution in the experiment cells at any time during the experiment run, as shown in Figure \ref{f:convoled}. The Gaussian mean (location of the Gaussian distribution peak) and standard deviations of the velocity distribution measured on free-flying aggregates lie at 1.84$_{-0.21}^{+0.30}v_{\textrm{max}}$. This value of the mean aggregate velocity will be used in the further data analysis.

\subsection{Number density of free-flying aggregates}
\label{s:n}
The data processing and analysis steps are described in Appendix~\ref{a:analysis}. This data analysis showed that the background greyscale of the averaged recorded frame provides information on the number of free-flying aggregates in the inner cell volume at each moment. In the following, this is quantified to determine the number density of free-flying aggregates during the different phases of the experiment.

The background greyscale value of each averaged frame was determined as being the maximum of the normalized histogram of this frame (see Figure \ref{f:analysis_gro}c., solid line). When no aggregates were flying in the cell volume, the background had the brightest possible value which is unity in the normalized greyscales shown in Figure \ref{f:analysis_gro}c. When all aggregates were free-flying, like at the very beginning of the experiment (see Figure~\ref{f:analysis_gro}c., dashed red line), the maximum normalized greyscale value was only around 0.7, due to the shadowing effect of the free-floating aggregates.

This normalized background greyscale value, $G$, is related to the optical depth of the aggregate system $\tau$, by $G$ = e$^{-\tau}$. The optical density is in turn related to the aggregate number density in the experiment cell volume $n$, by $\tau$ = $n~\sigma_{\textrm{mono}}~L$, where $\sigma_{\textrm{mono}}$ is the projected cross section of one monomer aggregate and $L$ is the depth of the experiment cell. Accordingly, the number density of free-flying aggregates is
\begin{equation}
n = -\frac{ln(G)}{\sigma_{\textrm{mono}}~L}\,.
\end{equation}
$G$ is measured as described above and $L$ is known. $\sigma_{\textrm{mono}}$ is calculated from the mean aggregate radius inside the experiment cell. The aggregate size distribution was measured in the laboratory (see Figure \ref{f:dust_distrib}, solid lines) and from the recorded experiment data during high velocity shaking phases (see Figure \ref{f:dust_distrib}, dashed lines). In these phases, individual aggregates cannot be tracked. However, the size distribution in still images can be measured. The data shows that the size distribution did not noticeably change during the flight: both cycle 1 and 3 final fast shaking phases show similar size distributions. As both the laboratory and the recorded size distributions are similar, the laboratory measurements can be used to asses $\sigma_{\textrm{mono}}$ = $\pi R_{\textrm{mono}}^2$, $R_{\textrm{mono}}$ being the measured aggregate radius (see Table~\ref{t:po} for measured radius values and their +/- 1$\sigma$ standard deviation).

\begin{figure}[tp]
  \begin{center}
  \includegraphics[width = 0.48\textwidth]{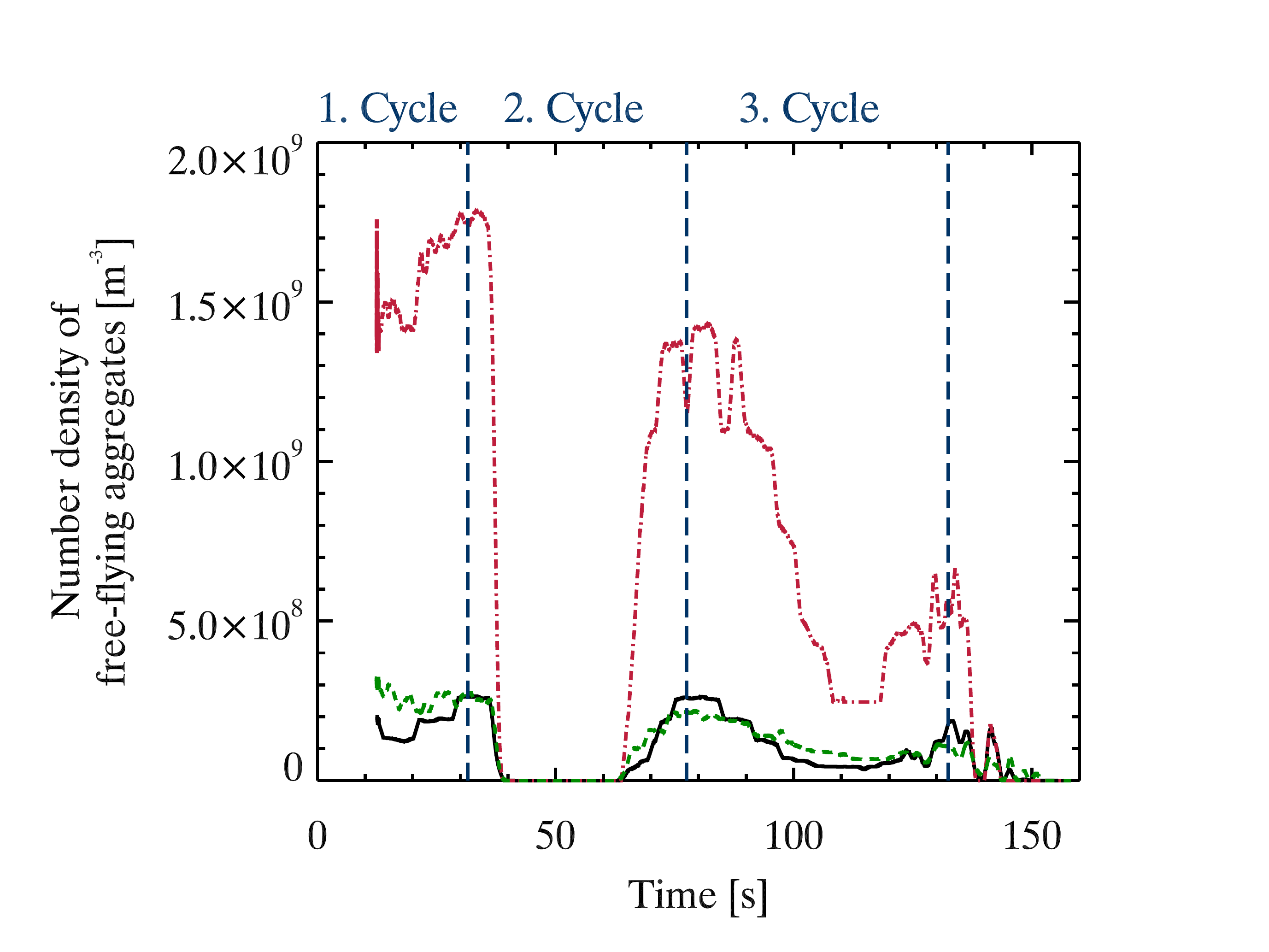}
 \caption{Retrieved number density of free-flying aggregates in the three cells of the SPACE experiment. The size distributions of large aggregates composed of polydisperse and monodisperse dust are shown as black solid and green dashed curves, respectively. The size distribution of small aggregates composed of monodisperse dust is shown as the red dash-dotted curve. The origin of time is at the start of data recording. The shaking cycles are indicated by dashed lines.}
 \label{f:n}
 \end{center}
\end{figure}

The resulting number density of free-flying aggregate is shown in Figure~\ref{f:n} as a function of time after start of the data recording. The slow shaking phase of the second cycle (see Figure~\ref{f:motor_profile}) can clearly be recognized at around 50~s, when no aggregates were free-flying in the volume. The two other slow shaking phases (cycles 1 and 3) were less pronounced at around 15 and 110~s, as both still had a high number of free-flying aggregates.

\subsection{Growth of clusters}
\label{s:area}
\begin{figure}[t]
  \begin{center}
  \includegraphics[width = 0.48\textwidth]{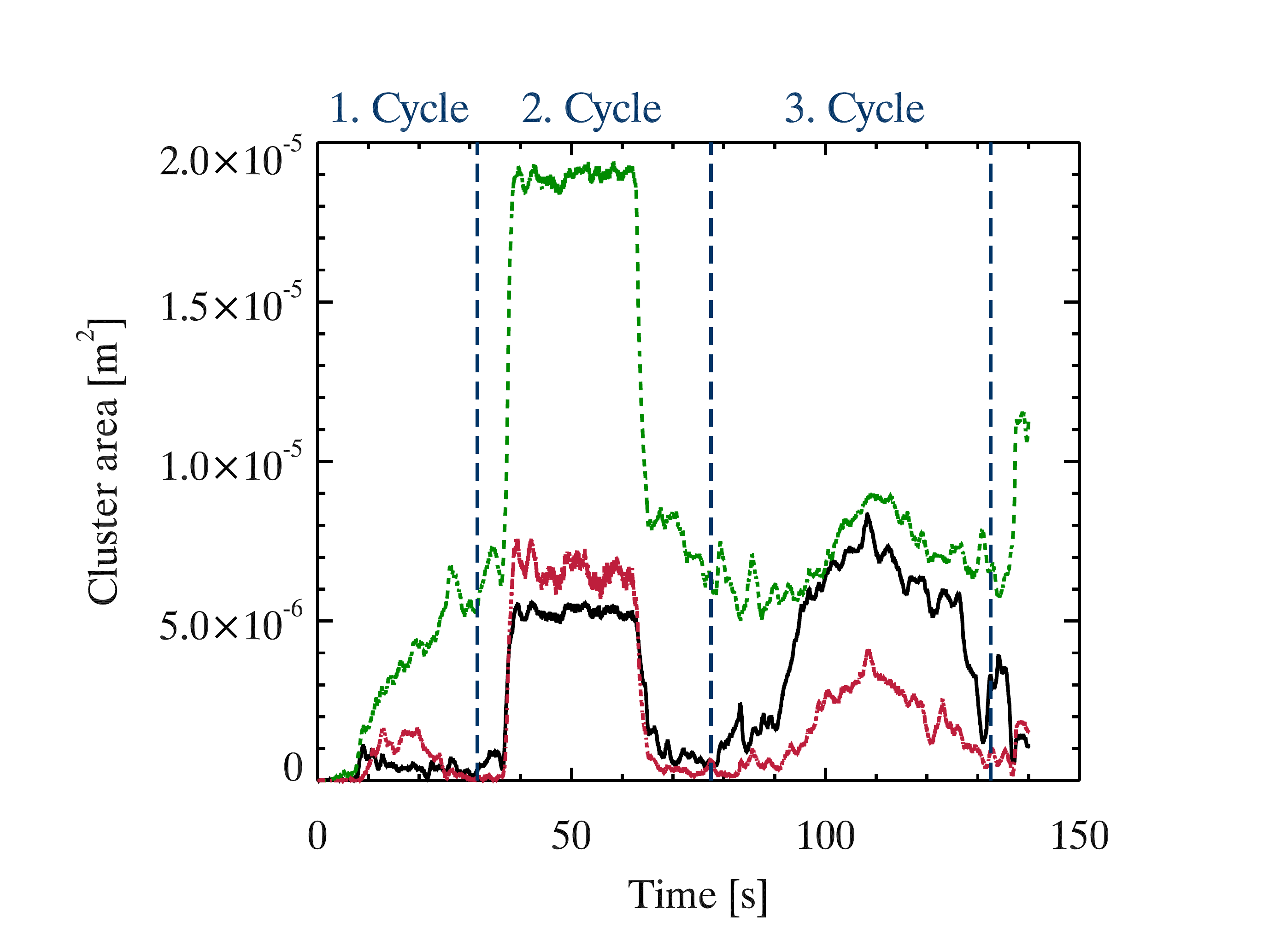}
 \caption{Cell wall surface area covered by clusters during the SPACE experiment for the three dust types. The large size distribution of aggregates composed of poly- and monodisperse dust are shown by the black solid and green dashed curve, respectively. The small size distribution of aggregates composed of monodisperse dust is shown by the red dash-dotted curve. The origin of time is at the start of data recording. The shaking cycles are indicated by dashed lines.}
 \label{f:area}
 \end{center}
\end{figure}

When averaging the corrected frames (see Appndix~\ref{a:analysis}), the clusters growing on the glass walls become visible (see Figure~\ref{f:analysis_gro}b.). Figure~\ref{f:area} shows the evolution of the total area covered by these clusters during the experiment run. During fast shaking phases, the clusters covered no or only small regions of the glass walls, while the maximum value of the cluster area was reached during the very slow shaking phase of cycle 2 at around 50~s after start of data recording.
The absolute value of the area covered by clusters growing on the cell walls appears to be higher for the large size distribution of aggregates composed of monodisperse dust than for the two other types of dust. However, this has no influence on the further results of our analysis as only derivative values were used (see below). As this difference is not reflected in the background greyscale analysis, it is likely that more growth seeds were created on the glass surfaces perpendicular to the line of sight, compared to the two other cells. Surfaces parallel to the line of sight, which do not get accounted for in the absolute value of wall cluster surfaces, probably gathered more of the flying aggregates.

\subsection{Aggregate pull-off forces}
\label{s:po}
By analysing the evolution of the clusters on the experiment cell walls during the fragmentation phases (17 to 32~s, 62 to 76~s and 112 to~132 s, see Figure~\ref{f:motor_profile}), it was possible to determine the normalized cluster fragmentation rate
\begin{equation}
f_r=\frac{1}{n}\frac{dn}{dt},
\end{equation}
with $n$ and $\frac{dn}{dt}$ being the number density of free-flying aggregates in the experiment cell and its time derivate, respectively. Aggregates were indeed detaching directly from the clusters, as no aggregate rolling on the cell walls was observed. Sticking tests of dust on the coated cell glass before the flight, as well as the very low sticking probability of the aggregates on the cell glass during the flight ($\sim$0.5\%, see Appendix~\ref{a:mc}), showed that the pull-off force of aggregates from the cell walls is lower than the pull-off force between two or more dust aggregates inside a cluster. In the high acceleration environment experienced by the aggregates during the suborbital flight, aggregates detaching from the walls instead of from a dust cluster would first be rolling on the wall before detaching. This behaviour was not observed.
\begin{figure}[t]
  \begin{center}
  \includegraphics[width = 0.48\textwidth]{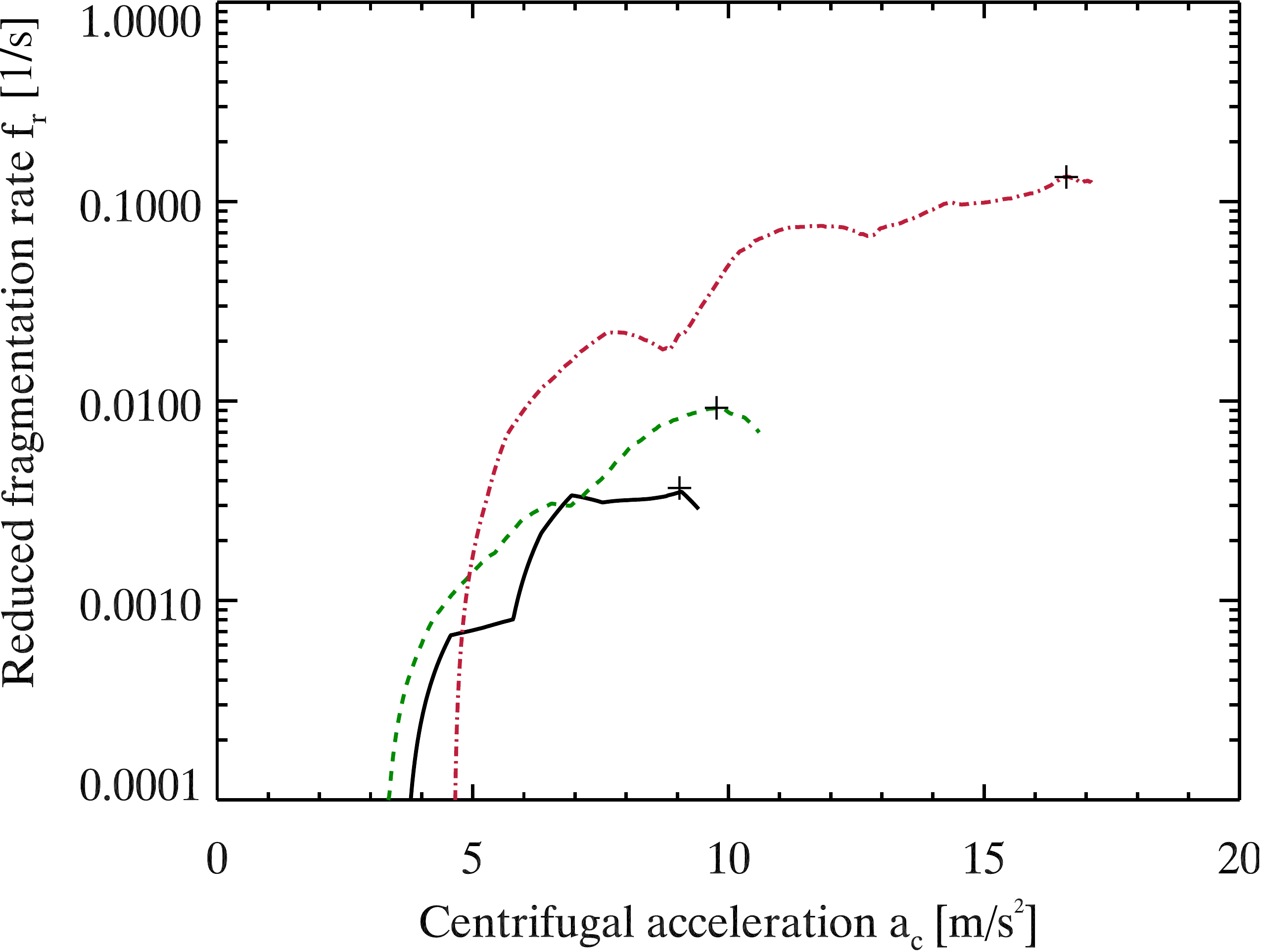}
 \caption{Normalized fragmentation rate of the clusters on the cell walls during the fragmentation phase of the second shaking cycle for the smaller and larger size distributions of aggregates composed of monodisperse SiO$_2$ (red dash-dotted for small aggregates and green dashed line for large aggregates, respectively) and the larger size distribution of aggregates composed of polydisperse SiO$_2$ (black solid line), plotted as a function of the centrifugal acceleration induced on the aggregates by the rotation of the cell wall. The curves are terminated at the acceleration where they reach the saturation of free-flying aggregates in the cell volume. The fragmentation rate maxima are shown by plus signs.}
 \label{f:frag_rate}
 \end{center}
\end{figure}

Figure~\ref{f:frag_rate} shows the measured normalized fragmentation rate as a function of the centrifugal acceleration induced by the cell wall rotation during the second shaking cycle. The curves are terminated at the acceleration where a saturation of free-flying aggregates in the cell volume was reached. The normalized fragmentation rate increases with increasing centrifugal acceleration, indicating that more and more aggregates were pulled off clusters on the wall. The maxima (marked in Figure~\ref{f:frag_rate}) of these reduced fragmentation rates indicate the accelerations at which most of the aggregates composing the clusters on the cell walls were pulled off their parent clusters and, thus, allow the determination of a pull-off force for each investigated aggregate size and type. For the smaller aggregates composed of monodisperse dust, the linear wall acceleration at maximum fragmentation was $a_{\textrm{c}}=16.6$~m~s$^{-2}$. The corresponding pull-off force is $F_{\textrm{po}}=ma_{\textrm{c}}=1.0_{-0.3}^{+1.7}\times10^{-8}$~N (see Table~\ref{t:po}), with $m$ being the mass of the aggregates pulled off the cluster. For the larger aggregates composed of mono- and polydisperse dust, the accelerations at maximum fragmentation were $a_{\textrm{c}}=9.8$~m~s$^{-2}$ and 9.1~m~s$^{-2}$, which corresponds to pull-off forces of $F_{\textrm{po}}=1.2_{-0.1}^{+0.1}\times10^{-7}$~N and 1.6$_{-0.4}^{+0.1}\times10^{-7}$~N, respectively.

\begin{figure}[tp]
  \begin{center}
  \includegraphics[width = 0.45\textwidth]{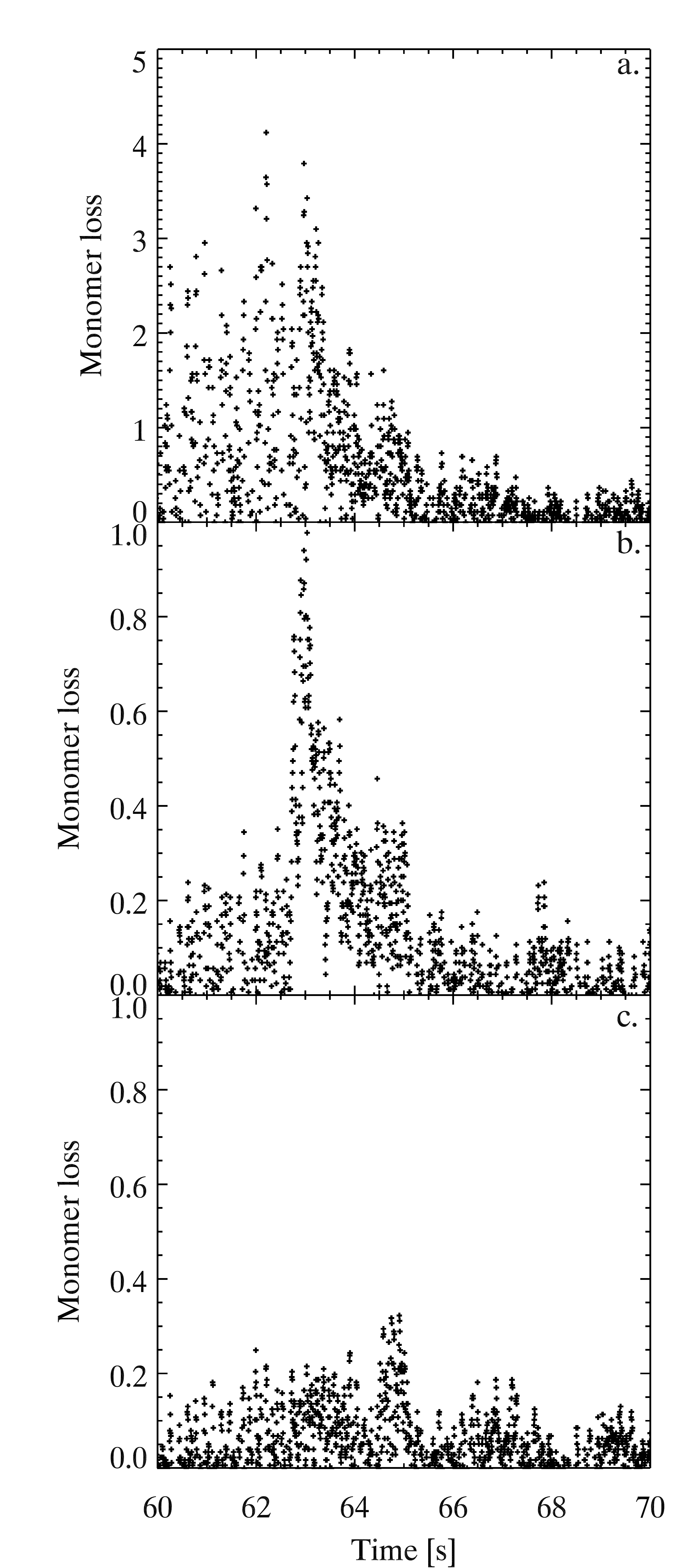}
 \caption{Monomer aggregate loss per cluster per frame (corresponding to 5.88~ms) for clusters on the cell walls during the fragmentation phase of the second shaking cycle for a.~the small size distribution of aggregates composed of monodisperse SiO$_2$, b.~the large size distribution of aggregates composed of monodisperse SiO$_2$, and c.~the large size distribution of aggregates composed of polydisperse SiO$_2$, respectively.}
 \label{f:mono_loss}
 \end{center}
\end{figure}

To verify whether the calculated pull-off forces are realistic, we determined the number of monomer aggregates that each cluster lost between two consecutive image frames during the fragmentation phase. The result of this investigation is shown in Figure \ref{f:mono_loss}. During the fragmentation phase, the shaking frequency of the experiment cells was increased over time, implying an increasing centrifugal acceleration level for the clusters on the wall. At centrifugal accelerations around the values for which the fragmentation rate of clusters reached its maximum ($\sim$63 s after start of data recording, see Figure~\ref{f:mono_loss}b.), the larger size distribution of aggregates composed of monodisperse dust displayes a clear peak of monomer losses. This peak is also visible, although less marked, for the other two dust types. The aggregates composed of polydisperse dust lost up to 0.4~monomers per cluster and time interval between two frames (5.88~ms) at $\sim$65~s (Figure~\ref{f:mono_loss}c.), meaning that not all clusters lost monomers at the same time. The small aggregates composed of monodisperse SiO$_2$, however, lost up to about 4~monomers per cluster per frame at $\sim$62~s (Figure~\ref{f:mono_loss}a.). This indicates that the small aggregates might have been pulled off the clusters in groups of four monomers instead of individually. The corrected aggregate radius and pull-off forces are listed in the last row of Table~\ref{t:po}. Since the aggregates pulled off the clusters were four times larger in surface at identical acceleration, the force required to pull them off was about one order of magnitude higher than for single aggregates.
\begin{table*}[t]
  \centering
  \caption{Accelerations at maximum fragmentation rate and pull-off forces $F_{\textrm{po}}$ for the three investigated types of dust. The corresponding tensile strengths $T_{\textrm{s}}=F_{\textrm{po}}/(\pi r_{\textrm{agg}}^2)$ are listed in the last column (see Section~\ref{s:tensile_strength}). The listed error values are implied by the +/- 1$\sigma$ standard deviation of the aggregate size distributions.}
    \begin{tabular}{|c|>{\centering\arraybackslash}m{1in}|>{\centering\arraybackslash}m{1.5in}|>{\centering\arraybackslash}m{0.75in}||>{\centering\arraybackslash}p{1.0in}|}
    \hline
    \multicolumn{1}{|c|}{\textbf{Aggregate type}} & \textbf{Mass averaged monomer aggregate radius $r_{\textrm{agg}}$ [$\mu$m]} & \textbf{Acceleration at maximum fragmentation rate [m~s$^{-2}$]} & \textbf{Pull-off force $F_{\textrm{po}}$ [N]} & \textbf{Tensile strength $T_{\textrm{s}}$ [Pa]}\\
\hline
    Small aggregates of& \multirow{2}{*}{59$_{-37}^{+68}$} & \multirow{2}{*}{16.6} & \multirow{2}{*}{1.0$_{-0.3}^{+1.7}\times10^{-8}$} &\multirow{2}{*}{1.0$_{-0.6}^{+1.1}$}\\
    monodisperse dust & & & &\\ \hline
    Large aggregates of &  \multirow{2}{*}{160$_{-62}^{+77}$} & \multirow{2}{*}{9.8}  & \multirow{2}{*}{1.2$_{-0.1}^{+0.1}\times10^{-7}$} & \multirow{2}{*}{1.6$_{-0.6}^{+0.7}$} \\
    monodisperse dust & & & & \\ \hline
    Large aggregates of& \multirow{2}{*}{163$_{-101}^{+72}$} & \multirow{2}{*}{9.1}  & \multirow{2}{*}{1.6$_{-0.4}^{+0.1}\times10^{-7}$} &\multirow{2}{*}{1.9$_{-1.2}^{+0.8}$}\\
    polydisperse dust & & & &\\ \hhline{|=|=|=|=||=|}
    Agglomerates of 4 aggregates & \multirow{2}{*}{118$_{-74}^{+137}$} & \multirow{2}{*}{16.6} & \multirow{2}{*}{8.4$_{-2.1}^{+13.2}\times10^{-8}$}&\multirow{2}{*}{1.9$_{-1.2}^{+2.2}$} \\
    of monodisperse dust & & & &\\ \hline
    \end{tabular}
  \label{t:po}
\end{table*}

\subsection{Sticking probability}
\label{s:sticking_prob}

\begin{figure}[t]
  \begin{center}
  \includegraphics[width = 0.48\textwidth]{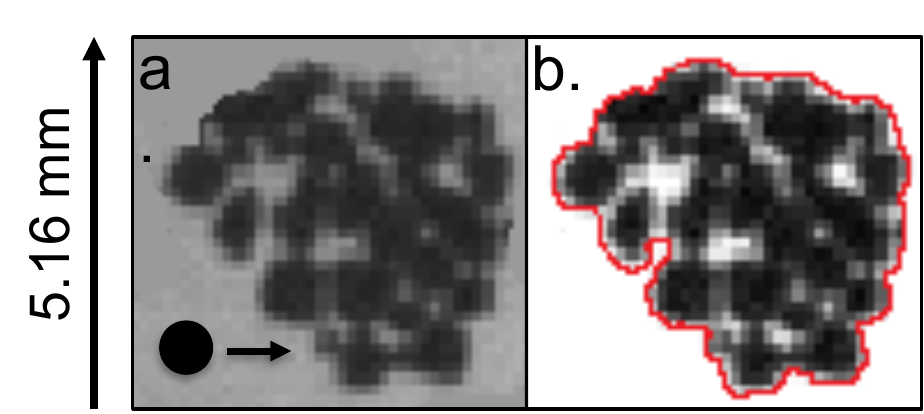}
 \caption{Typical wall cluster during the experiment from the large size distribution of aggregates composed of monodisperse dust. a. Original greyscale image. b. Contrast enhanced image. The red line in b. indicates the delimitation of the total cluster area. The packing density $\phi_{\textrm{p}}$~=~0.91 is determined by dividing the dark area by the area enclosed by the red line (see Section \ref{s:cluster_morphology} for more details on the cluster morphology).}
 \label{f:phip}
 \end{center}
\end{figure}

The analysis of the aggregate behaviour for increasing centrifugal accelerations shows that for accelerations under $\sim$4~m~s$^{-2}$, fragmentation plays no role in the evolution of clusters of aggregates (see Figure~\ref{f:frag_rate}). During the second shaking cycle, aggregate sticking and cluster formation was observed at cell wall accelerations below 4~m~s$^{-2}$ (see Figure~\ref{f:motor_profile}). This means that the relatively high acceleration environment due to the shaking of the experiment cells did not influence the sticking behaviour of the aggregates. Its effects were only noticeable in the morphology of the formed clusters (see more details in Section~\ref{s:cluster_morphology} and Figure~\ref{f:phip}). We therefore proceeded with the analysis of the sticking behaviour of the studied dust aggregates and measured their sticking probability as described in this section.\\
The number density profile determined in Section~\ref{s:n} can be used to derive the growth rate $dn/dt$ of clusters during shaking velocity ramp-down phases (6 to 15~s, 35 to 46~s and 85 to 112~s, see Figure~\ref{f:motor_profile}), which in turn can be used to calculate the sticking probability $\beta$ of the coagulating system of aggregates.\\
Smoluchowski's equation for particle coagulation can be adapted to the assumption that the growing clusters in the SPACE experiment were composed of a number $i$ of single monomer aggregates \citep[see][]{smoluchowski1916PZ,blum2006AP}:
\begin{equation}
\label{e:smolu}
\begin{split}
\frac{\partial n(i, t)}{\partial t}=\frac{1}{2}\sum_{j=1}^{i-1} K_a(j, i-j)n(j, t)n(i-j, t)\\
-n(i,t)\sum_{j=1}^{\infty}K_a(j,i)n(j,t)
\end{split}
\end{equation}
Here, $n(i,t)$ is the number density of clusters composed of $i$ monomer aggregates at a time $t$ and $K_a$ is the collision kernel for ballistic collisions between aggregates/clusters composed of $i$ and $j$ monomer aggregates, respectively. In this equation, the first term on the right-hand side accounts for the creation of clusters composed of $i$ monomers by differently-sized aggregates, while the second term accounts for the depletion of clusters composed of $i$ monomers by sticking collisions with other aggregates. The collisions taking place in the SPACE experiment cells during growth phases were mainly between monomer aggregates ($j=1$) and clusters growing on the cell walls, thus depleting the inner cell volume of free-flying aggregates. Assuming that all clusters contain the same number of monomer aggregates at any given time, this implies that the first term on the right-hand side of Equation~\ref{e:smolu} vanishes.

In the second term on the right-hand side, the collision kernel is defined as
\begin{equation}
K_a(j,i)=\beta(j,i;v)~v(j,i)~\sigma(j,i)
\end{equation}
Here, $v(j,i)$ is the relative velocity between the clusters and the monomers, $\sigma(j,i)$ the collision cross section between the two, and $\beta(j,i;v)$ the sticking probability of their collision, respectively. For the collisions considered here, the sticking probability and relative velocity are considered to be the same for all aggregates at each moment. Hence, Equation \ref{e:smolu} becomes
\begin{equation}
\label{e:smolu_bis}
\frac{dn(t)}{dt}=-n(t)n'(t)\beta(t)\sigma_{\textrm{cross}}(t)v(t),
\end{equation}
where $n(t)$ is the number density of free-flying aggregates in the cell volume at time $t$ determined in Section~\ref{s:n}, $dn(t)/dt$ is the derivative of the number density of aggregates incorporated in clusters on the cell walls determined in Section~\ref{s:area}, $\beta(t)$ and $v(t)$ are the sticking probability and relative velocity of the collisions between the monomer aggregates and the clusters on the cell walls, $\sigma_{\textrm{cross}}(t)$ is the mean collision cross section between a free-flying aggregate and a cluster on the wall, and $n'(t)$ is the number density of clusters on the cell walls.
The relative velocity $v$ between the free-flying monomer aggregates and the rotating clusters on the cell walls has been determined in Section~\ref{s:velocity}. The mean collision cross section $\sigma_{\textrm{cross}}$ was calculated at each moment $t$ of the experiment run by averaging over the collision cross sections of all clusters on the cell walls,
\begin{equation}
\begin{split}
\sigma_{\textrm{cross}}(t) = \frac{1}{N(t)}\sum_{k = 0}^ {N(t)}\sigma_{cross,k}\\
 = \frac{1}{N(t)}\sum_{k = 0}^ {N(t)}2R_k(t)\cdot 2r\\
 = \frac{4r}{N(t)}\sum_{k = 0}^ {N(t)}R_k(t)
\end{split}
\label{e:cs}
\end{equation}
\begin{figure}[t]
  \begin{center}
  \includegraphics[width = 0.48\textwidth]{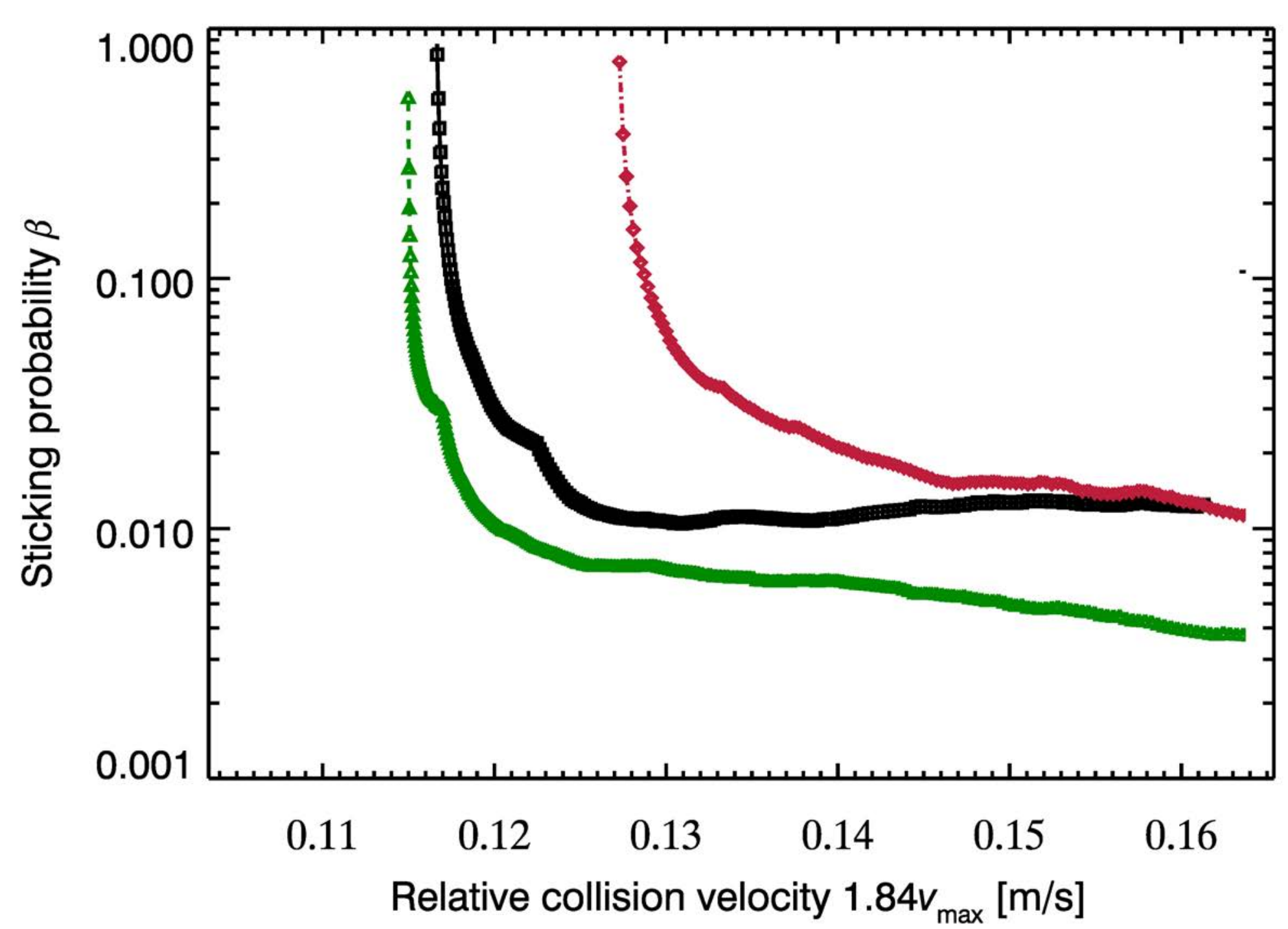}
 \caption{Sticking probability of the three types of dust aggregates investigated in the SPACE experiment plotted as a function of the statistical relative collision velocity computed in Section~\ref{s:velocity} (1.84$v_{\textrm{max}}$). The large size distribution of aggregates composed of poly- and monodisperse dust are shown by the black solid curve with squares and the green dashed curve with triangles, and the small size distribution of aggregates composed of monodisperse dust is shown by the red dash-dotted curve with diamonds.}
 \label{f:beta}
 \end{center}
\end{figure}
where $N(t) = n'(t)V$ is the total number of clusters on the cell walls, $V$ being the cell volume, and $R_k$ and $r$ are the radii of the wall cluster $k$ and a monomer aggregate, respectively. In equation~\ref{e:cs}, the collision cross section $\sigma_{\textrm{cross}}$ is taken as $2R_k(t)\cdot 2r$. This assumes that the clusters on the cell walls have a thickness of one monomer aggregate and that the monomer aggregates approach them perpendicular to the observation direction, due to their motion induced by wall collisions (this collision configuration is shown by an artificially drawn incoming aggregate in the bottom left corner of Figure~\ref{f:phip}a). To support this assumption, a typical wall cluster was chosen in the cell containing the larger size distribution of aggregates composed of monodisperse dust (see Figure~\ref{f:phip}). In this cluster, the constituting monomer aggregates can very well be distinguished. This indicates that the cluster is very thin (most likely only one layer thick) and its packing density $\phi_{\textrm{p}}$ can be determined as the ratio between the area covered by aggregates to the total area of the cluster. For this cluster, the packing density is $\phi_{\textrm{p}}$~=~0.91. As the densest packing of circles on a surface is reached for hexagonal packing, where $\phi_{\textrm{p}}$~=~0.9069, it seems an appropriate approximation to assume that the clusters growing on the wall during the SPACE experiment possess a thickness of about one monomer (see Section \ref{s:cluster_morphology} for more details on the cluster morphology). Thus, the quantity
\begin{equation}
\sigma_{\textrm{cross}}(t)n'(t)=\frac{4r}{V}\cdot\sum_{k = 0}^ {N(t)}R_k(t)
\end{equation}
can be measured from the analysis of clusters growing on the wall during the SPACE experiment (see Section~\ref{s:area}).

Finally, the sticking probability can be calculated as
\begin{equation}
\beta(t)=-\frac{1}{n(t)}\cdot\frac{dn(t)}{dt}\cdot\frac{1}{\sigma_{\textrm{cross}}(t)n'(t)v(t)} \,.
\label{e:beta}
\end{equation}
All the quantities on the right hand side of this equation are measured at each frame during the experiment run. The resulting sticking probability between the monomer aggregates and the clusters on the cell walls calculated with Equation~\ref{e:beta} is shown in Figure~\ref{f:beta} as a function of the mean collision velocity and for all three aggregate types. During the growth phases of cycles 1 and 3, the sticking probability was very close to $\beta = 0$ because the wall speed was $\sim$13~cm~s$^{-1}$ (see Figure~\ref{f:motor_profile}) so that the mean collision speed of 24~cm~s$^{-1}$ was much larger than the velocity range for sticking (see Figure~\ref{f:beta}). Accordingly, only the sticking probability during the growth phase of cycle 2 is shown. The curves that can be distinguished growing in sticking probability with decreasing impact velocity are for the three different dust types during the growth phase of cycle 2. Below a certain velocity, the sticking probability of the dust aggregates rises very steeply and approaches $\beta = 1$. At this point, no free-flying aggregates are left in the cell volume. This velocity was determined to be the maximum velocity at which all aggregates are incorporated in clusters, leaving none free-flying in the cell volume. Table~\ref{t:vc} lists the determined values for all three investigated types of aggregates. Both, the larger distributions of aggregates composed of mono- and polydisperse SiO$_2$, showed perfect sticking at similar velocities of 11.5$_{-1.3}^{+1.9}$ and 11.7$_{-1.3}^{+1.9}$~cm~s$^{-1}$, respectively, while the smaller distribution of aggregates composed of monodisperse SiO$_2$ stuck perfectly at 12.7$_{-1.4}^{+2.1}$~cm~s$^{-1}$ (error values are implied by the +/- 1$\sigma$ standard deviation on the collision velocities, see Section~\ref{s:velocity}). The minimum velocity for which $\beta<$~0.05 was also determined. The smaller aggregates composed of monodisperse dust reached $\beta=0.05$ at a mean collision velocity of 13.4$_{-1.5}^{+2.2}$~cm~s$^{-1}$ and the larger aggregates composed of mono- and polydisperse dust reached $\beta=0.05$ at 12.2$_{-1.4}^{+2.1}$ and 12.6$_{-1.4}^{+2.1}$~cm~s$^{-1}$, respectively.

\begin{table}[t]
\caption{Perfect sticking velocities for all three aggregate types during the SPACE suborbital flight experiment. The velocities for which $\beta>0.05$ are also listed. The error values are implied by the +/- 1$\sigma$ standard deviation on the collision velocities (see Section~\ref{s:velocity} for more details)}
\begin{center}
\begin{tabular}{|c|>{\centering}p{0.95in}|>{\centering\arraybackslash}p{0.9in}|}
\hline
\textbf{Aggregate type} & \textbf{Velocity for perfect sticking [cm~s$^{-1}$] } & \textbf{Velocity above which \boldmath$\beta<$0.05 [cm~s$^{-1}$]} \\ \hline
Small aggregates of &  \multirow{2}{*}{12.7$_{-1.4}^{+2.1}$} &  \multirow{2}{*}{13.4$_{-1.5}^{+2.2}$}      \\
monodisperse dust &&\\ \hline
Large aggregates of   &  \multirow{2}{*}{11.5$_{-1.3}^{+1.9}$}   &  \multirow{2}{*}{12.2$_{-1.4}^{+2.1}$}   \\
monodisperse dust &&\\ \hline
Large aggregates of   & \multirow{2}{*}{11.7$_{-1.3}^{+1.9}$}  &  \multirow{2}{*}{12.6$_{-1.4}^{+2.1}$} \\
polydisperse dust &&\\\hline
\end{tabular}
\end{center}
\label{t:vc}
\end{table}

\section{Discussion}
\label{s:discussion}

During the suborbital flight, the dust aggregates inside the SPACE experiment cells were subjected to residual accelerations and potentially also to magnetic or electrostatic effects. We investigated the influence of these forces on the dust aggregates and present those in Appendix~\ref{a:other_influences}. The result of this analysis is that aggregate motion induced by rocket residual accelerations, magnetic, or electrostatic effects can all be neglected compared to the accelerations induced by the cell shaking. \\
In this Section, we will start by comparing the present experiment results to previous work in Section~\ref{s:compare}, discuss the morphology of the formed aggregate clusters in Section~\ref{s:cluster_morphology} and derive the tensile strength of clusters and the surface energy of their aggregates in Section~\ref{s:cluster_properties}. Finally,  conclusions will be derived for the behaviour of dust aggregates and clusters in protoplanetary disks in Section~\ref{s:ppd}.

\subsection{Comparing the SPACE results with previous work}
\label{s:compare}

The aggregate collisions investigated in this paper are different from the ones observed in previous work performed by \citet{weidling_et_al2012Icarus} and \citet{kothe_et_al2013Icarus}. In the present work, aggregates did not collide while free-flying. Instead, one of the collision partners was attached to the glass cell wall and moving in a circular motion. Unlike in \citet{weidling_et_al2012Icarus} and \citet{kothe_et_al2013Icarus}, this collision configuration prevented the formation of fractal clusters and collision energy did not dissipate through cluster restructuration or vibration. Instead, the relatively high acceleration environment at which the dust aggregates evolved during the SPACE experiment run imposed restructuring of the growing aggregates into compact mono-layered clusters (see Section~\ref{s:cluster_morphology} below). The sticking and fragmenting behaviour of the aggregates, however, was not influenced as we showed in Sections \ref{s:po} and~\ref{s:sticking_prob}. We could therefore deduce the aggregate and cluster properties described below in Section~\ref{s:cluster_properties}.


\begin{figure}[t]
  \begin{center}
  \includegraphics[width=0.45\textwidth]{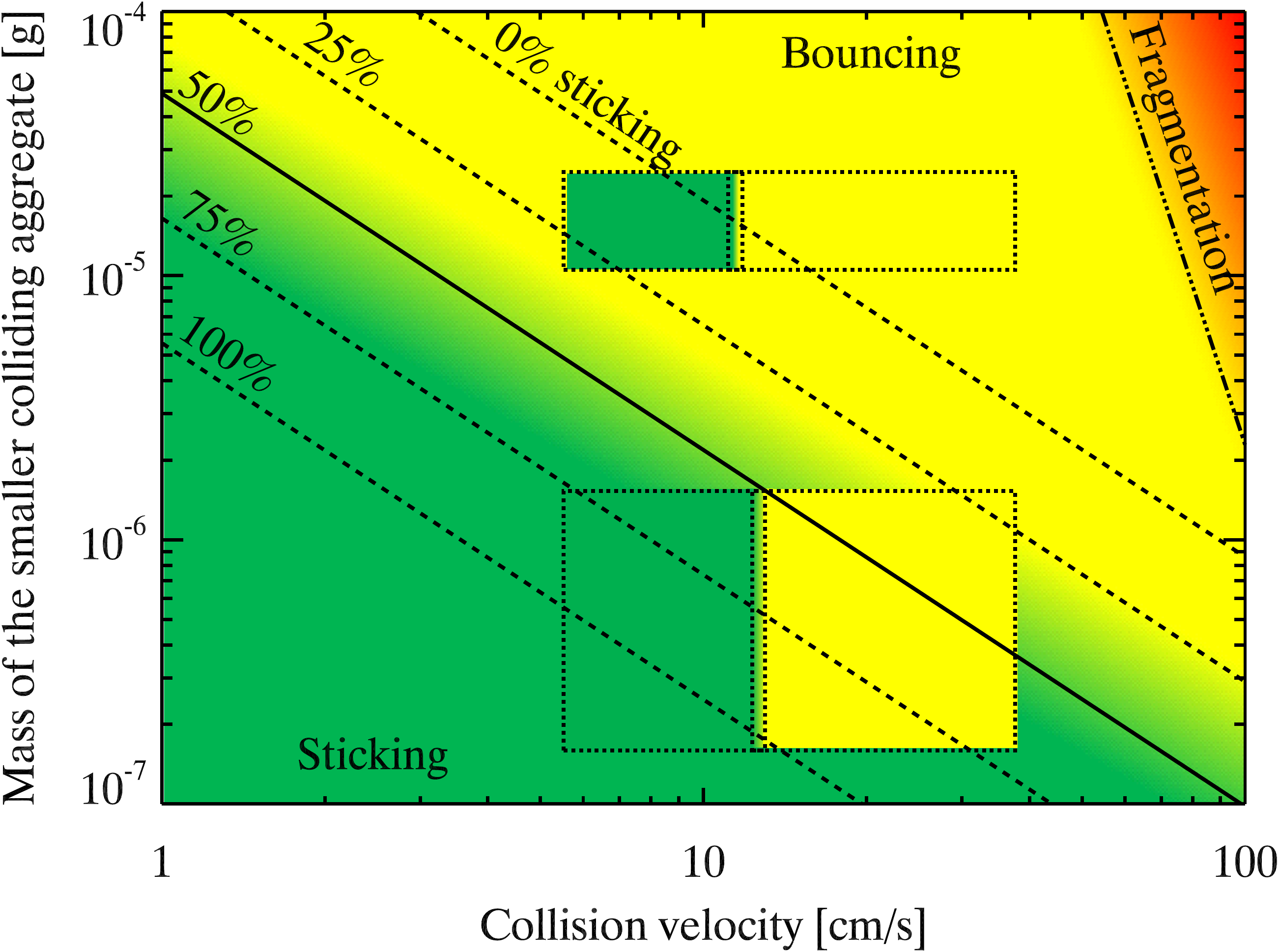}
 \caption{Flight collision results (dotted boxes) according to the aggregate size distributions (box heights) and their mean collision velocities during the experiment (box widths). As the larger aggregates composed of mono- and polydisperse SiO$_2$ grains had very similar sticking properties, only the results for the aggregates composed of monodisperse SiO$_2$ are plotted (upper box). The background colours correspond to the dust collision model developed by \citet{guettler_et_al2010A&A} with updates by \citet{kothe_et_al2013Icarus}. Green colour represents sticking collisions, yellow bouncing, and red  fragmentation, respectively. The lines represent the transition zones between sticking and bouncing (solid line for a 50\% sticking probability, dashed lines for 0, 25, 75 and 100~\% sticking probabilities) as well as the transition between bouncing and fragmentation (dash-dotted line for the onset of fragmentation), computed by \citet{kothe_et_al2014prep}.  Mind that the collision model by \citet{guettler_et_al2010A&A} and \citet{kothe_et_al2013Icarus} was developed for collisions between equal-mass aggregates, whereas the results presented here are dominated by single aggregates colliding with more massive clusters.}
 \label{f:res_rex}
 \end{center}
\end{figure}

Figure \ref{f:res_rex} compares the SPACE results with the dust collision model published by \citet{guettler_et_al2010A&A} with updates by \citet{kothe_et_al2013Icarus}. The collisions observed during the experiment run are represented by boxes with the box height given by their initial size distribution and the box width given by the range in mean relative collision velocity. The velocities for the onset of sticking (sticking probability $\beta=0.05$) and for perfect sticking ($\beta\simeq1$) listed in Table~\ref{t:vc} are added to the boxes (vertical dotted lines), with the green and yellow regions denoting sticking and bouncing, respectively. As the larger aggregates composed of mono- and polydisperse dust had very similar sticking velocities, only the results for the monodisperse aggregates are plotted (upper box). The dust collision model is displayed in the background, delimiting the regions of the parametre range where sticking (green), bouncing (yellow), and fragmentation (red), respectively, dominate for collisions between similar-sized dust aggregates.

It can be noted that many sticking collisions occur in parametre ranges where bouncing is predicted by the current model. One explanation for this unexpected collision behaviour is the fact that the observed events are not aggregate-aggregate collisions, for which the model was developed, but aggregate-cluster collisions between free-flying aggregates and clusters growing on the cell walls. The mass ratio between colliding aggregates during the SPACE experiment covered up to 4 orders of magnitude. At the highest mass ratio between target cluster and projectile aggregate of $\sim10^4$, only sticking collisions were observed. \citet{kothe_et_al2013Icarus} (see their Fig. 8) obaserved the enhanced sticking probability between clusters composed of a large number of aggregates, compared with the results of \citet{weidling_et_al2012Icarus} who analysed collisions between aggregates of similar mass that resulted in bouncing for more than 90~\% of the collisions at velocities between 0.2~and~50~cm~s$^{-1}$. However, unlike both \citet{kothe_et_al2013Icarus} and \citet{weidling_et_al2012Icarus} who analysed collisions between same-sized aggregates, the data presented here reveals the role of the mass ratio between the colliding aggregates.

In \citet{weidling_et_al2012Icarus} and \citet{kothe_et_al2013Icarus}, the coexistence of sticking and bouncing collisions in an extended region of the parametre space was observed. However, the rocket flight data display a very sharp transition from one regime to the other.  This can be attributed to the statistical nature of the measurement: as the collision outcome is averaged over a high number of collisions, the transition appears sharper than for the individual collisions observed in the drop tower experiments.

\subsection{Cluster morphology} 
\label{s:cluster_morphology}

The flat cluster structures found in our experiments (see Figure \ref{f:phip}) is due to an immediate restructuring of the clusters following a sticking collision. Due to the relatively high acceleration environment ($\stackrel{>}{\sim}$~1~m~s$^{-2}$) during the agglomeration phases (see Figure~\ref{f:motor_profile}), the centrifugal force experienced by the newly-acquired dust aggregate can potentially force individual aggregates to roll radially outward until no further rolling motion is possible. To evaluate whether this was the case in the SPACE experiment, we compare the centrifugal force with the rolling-friction force that counteracts any rolling motion. The latter was observed by \citet{beitz_et_al2012Icarus}  (their Figure 4), who witnessed the rolling of a dust aggregate above another under microgravity conditions and found that the initial velocity v$\sim$2~cm~s$^{-1}$ was reduced to zero within a length of h$\sim$4~mm. From these data we derive a rolling-friction acceleration of a$\sim v^2/2h$=0.1~m~s$^{-2}$, which is below the minimum centrifugal acceleration of ~1~m~s$^{-2}$. Thus, we expect all aggregates to roll (but not to detach!) after sticking to the cluster. 

The flattened, quasi-two-dimensional nature of the clusters, for which all constituent aggregates are in contact with the wall, considerably enhances the rigidity of the cluster. This, in turn, has consequences for the impinging dust aggregates. In contrast to the cluster-cluster collisions observed by \citet{kothe_et_al2013Icarus} in which the weakly bound clusters could dissipate a much larger share of the impact energy, an impacting aggregate in our experimental configuration meets an aggregate that is attached to a stiff cluster. Hence, the physical situation is similar to an aggregate-aggregate collision, in which the impact energy is dissipated within the two aggregates in contact, with the exception that one of the aggregates has infinite mass \citep[see also][for more details]{brisset2014_diss}. Therefore, the experiments performed here under low residual acceleration conditions are capable of deriving the sticking threshold of dust aggregates in mutual collisions but not the morphology of the clusters that would form under a perfect zero-gravity condition.

From numerical and experimental work by e.g. \citet{blum_et_al2000PRL, krause_and_blum2004PRL, blum_et_al2006PRL, blum2006AP, kothe_et_al2013Icarus}, we know that growth through aggregate-aggregate and aggregate-cluster collisions in the sticking regime leads to very fluffy and fractal cluster structures. However, the collision environment of the protoplanetary nebula is not restricted to the hit-and-stick regime and can also lead to the formation of clusters that were rendered compact and stiff by bouncing collisions \citep{weidling_et_al2009ApJ, zsom_et_al2010AA, guettler_et_al2010A&A}. The structure of the clusters formed in the SPACE experiment is therefore, as well as the fractal aggregates observed in \citet{kothe_et_al2013Icarus}, representative of a particle population of the protoplanetary nebula.

\subsection{Properties of clusters composed of sub-mm-sized aggregates}
\label{s:cluster_properties}

Here, the clusters built during the growth phases of the SPACE experiment will be investigated and some of their properties derived. Their inner cohesion is studied by investigating the tensile strength and the surface energy of their sub-mm-sized constituents.

\paragraph{Tensile strengths of aggregate clusters}
\label{s:tensile_strength}

The pull-off forces required to detach a monomer aggregate from a wall cluster built during the slow shaking phase of the suborbital flight experiment are listed in Table~\ref{t:po}. From these pull-off forces, the corresponding tensile strengths of clusters composed of monomer aggregates could be determined through $T=F_{\textrm{po}}/\sigma_{\textrm{agg}}$, with $\sigma_{\textrm{agg}}$ being the cross section of the detaching aggregate. We obtained 2.0$_{-0.6}^{+0.8}$~Pa and 2.1$_{-1.0}^{+1.1}$~Pa for the larger size distributions of aggregates composed of monodisperse and polydisperse dust, respectively. The tensile strength of agglomerates of four monomer aggregates composed of monodisperse dust is 2.7$_{-1.4}^{+1.3}$~Pa (see Section~\ref{s:po} for details).

\begin{figure}[t]
  \begin{center}
  \includegraphics[width = 0.5\textwidth]{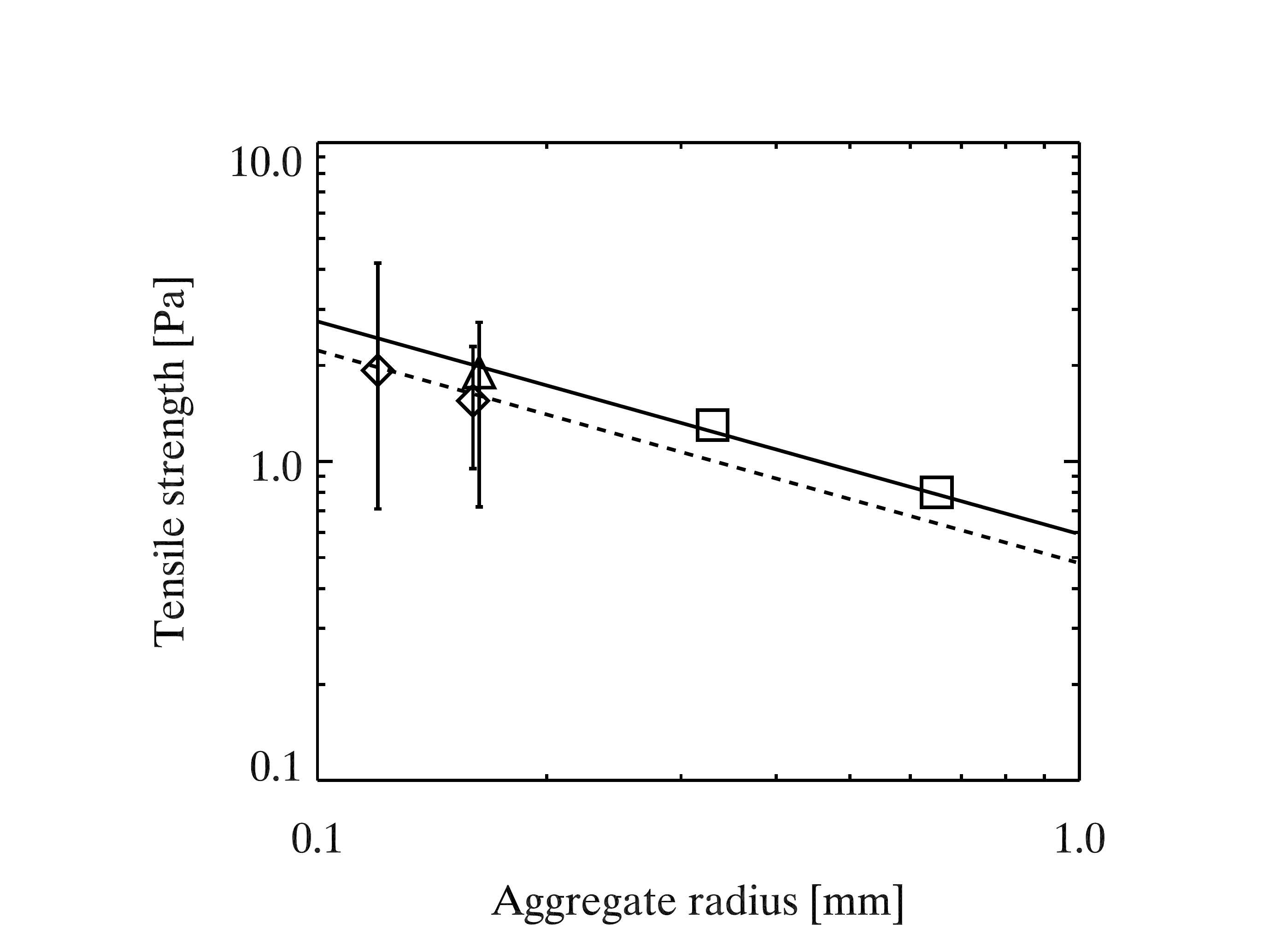}
 \caption{Tensile strength of aggregates in clusters investigated in the SPACE experiment for aggregates composed of mono- and polydisperse dust (diamonds and triangle, respectively) as well as of mm-sized dust aggregates measured by \citet{blum_et_al2014Icarus} (squares) in comparison to the model prediction by \citet{skorov_and_blum2012Icarus}. The model is applied for aggregate filling factors of 0.3 (dashed line) and 0.37 (solid line), respectively.}
 \label{f:ts}
 \end{center}
\end{figure}

The three values of the tensile strength are plotted in Figure~\ref{f:ts} as a function of the size of the detaching aggregates (diamonds and triangle for aggregates composed of mono- and polydisperse SiO$_2$, respectively). The lines show the predicted values of the tensile strength of aggregates following the model by \citet{skorov_and_blum2012Icarus}, i.e.
\begin{equation}
T = T_1\phi\left(\frac{r}{1\textrm{ mm}}\right)^{-\frac{2}{3}} ,
\end{equation}
where $r$ and $\phi$ are the radius and the packing density of the aggregate. \citet{skorov_and_blum2012Icarus} derived for the parametre $T_1$ = 1.6 Pa. In Figure \ref{f:ts}, the dashed line is the model tensile strength computed for an aggregate packing density of $\phi =0.3$ as used by \citet{skorov_and_blum2012Icarus} and the solid line for a packing density of $\phi =0.37$, the value assumed for the aggregates used in the SPACE experiment (see Section~\ref{s:dust}). Keeping in mind that the model predictions have no free parametre, the data from the SPACE experiment are in very good agreement with the predictions by \citet{skorov_and_blum2012Icarus}.
Furthermore, \citet{blum_et_al2014Icarus} also measured the tensile strength of SiO$_2$ aggregate clusters, for aggregate diametres of 0.6 and 1.2 mm, respectively. Their results are plotted as squares in Figure \ref{f:ts} and are in excellent agreement with the model by \citet{skorov_and_blum2012Icarus} and the SPACE data. The concurrence of the theoretical model by \citet{skorov_and_blum2012Icarus} with the results of two independent experimental methods is a strong support of the validity of the model for the derivation of the tensile strength of aggregate clusters. Please mind that the internal tensile strength of the dust-aggregate material with an individual aggregate is typically on the order of a few kPa \citep{blum_et_al2006ApJ}. The much lower tensile strength of adhering dust aggregates results from the relatively small contact area between the aggregates.

\paragraph{Estimation of the surface energy of macroscopic aggregates}

\begin{figure}[t]
  \begin{center}
  \includegraphics[width = 0.48\textwidth]{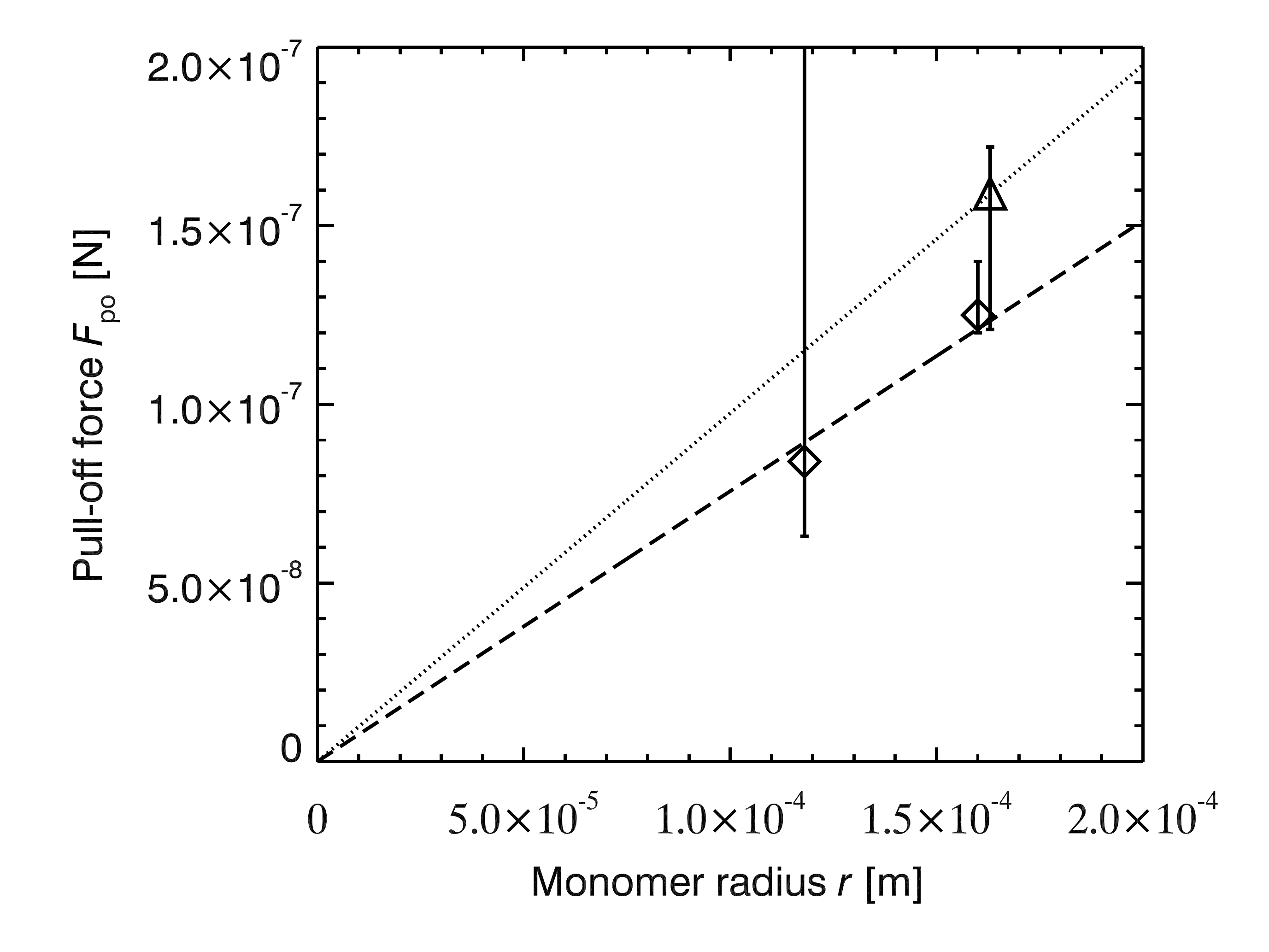}
 \caption{Aggregate pull-off forces as a function of their size for the two types of dust investigated in the SPACE suborbital flight experiment, i.e. monodisperse (diamonds) and polydisperse SiO$_2$ (triangle). Linear fits through the data and the origin are also shown, with slopes of 7.6$\times10^{-4}$ (dashed line) and 9.8$\times10^{-4}$ (dotted line) for the two dust types, respectively. The smaller aggregates composed of monodisperse dust are assumed to detach in clumps of four aggregates.}
 \label{f:gamma}
 \end{center}
\end{figure}
The pull-off forces listed in Table~\ref{t:po} are plotted in Figure~\ref{f:gamma} as a function of the monomer aggregate size. For the smaller size distribution of aggregates composed of monodisperse dust, the detaching of agglomerates of four aggregates was assumed, as described in Section~\ref{s:po}. The Johnson-Kendall-Roberts theory \citep{johnson_et_al1971PRSL} for the contact between soft spheres predicts a pull-off force
\begin{equation}
F_{\textrm{po}} = 3\pi\gamma r
\end{equation}
of one sphere from another as a function of the radius $r$. Here, $\gamma$ is the surface energy of the material that forms the contact between the two spheres. Although this applies to perfect macroscopic soft spheres, the above relationship has been observed to be valid for $\mu$m-sized SiO$_2$ particles as well \citep{heim_et_al1999PRL}. If this relationship is assumed to be valid for the monomer aggregates investigated by the SPACE experiment (mean diametres of $\sim$120 and $\sim$330 $\mu$m), an effective surface energy $\gamma_{\textrm{eff}}$ must be introduced, as done in \citet{weidling_et_al2012Icarus}. Following their method, the effective surface energy can be scaled to the surface energy $\gamma$ of the constituent micrometre-sized monomer particles via the aggregate filling factor $\phi$ and the Hertz factor $a^2/a_0^2$, where $a_0$ is the radius of a monomer particle and $a$ is the radius of the contact surface between two such monomer particles. This scaled effective surface energy can be written as
\begin{equation}
 \gamma_{\textrm{eff}} = 2\,N\,\gamma\,\phi\,\frac{a^2}{a_0^2}
\end{equation}
where $N$ is the number of connections from the separating aggregate to the cluster aggregates. The factor 2 accounts for the two sticking aggregates each having a surface energy of $\gamma$. As derived by \citet{johnson_et_al1971PRSL} \citep[see][for more details]{weidling_et_al2012Icarus},
\begin{equation}
\frac{a^2}{a_0^2} = \left(\frac{9\pi\gamma(1-\nu^2)}{a_0E_0}\right)^\frac{2}{3},
\end{equation}
where $E_0$ and $\nu$ are the monomer-particles' Young's modulus and Poisson number, respectively. Therefore,
\begin{equation}
\gamma = \left(\frac{\gamma_{\textrm{eff}}}{2\,N\,\phi\left(\frac{9\pi(1-\nu^2)}{a_0E_0}\right)^\frac{2}{3}}\right)^\frac{3}{5}.
\end{equation}
For the SiO$_2$ dust used in the SPACE experiment, we assume $\nu = 0.17$, $E_0=5.4\times10^{10}$~Pa, $a_0=7.6\times10^{-7}$~m and $\phi=0.37$. The number of connections from the separating aggregate to the cluster aggregates was determined from a typical cluster (Figure \ref{f:phip}) to be 2.5. In this cluster, each aggregate on the rim had between 2 and 3 neighbours.
\begin{table}[t]
  \centering
  \caption{Effective surface energies for the dust investigated in the SPACE rocket flight experiment.}
    \begin{tabular}{|>{\centering}p{0.7in}|>{\centering}p{0.7in}|>{\centering}p{0.7in}|>{\centering\arraybackslash}p{0.7in}|}
    \hline
    \textbf{Aggregate type} & \textbf{Measured effective surface energy [J~m$^{-2}$]} & \textbf{Measured effective surface energy for one contact [J~m$^{-2}$]} & \textbf{Measured surface energy scaled down to a monomer particle [J~m$^{-2}$]}\\ \hline
    Aggregates composed of& 8.0$\times10^{-5}$ & 1.6$\times10^{-5}$ &1.7$\times10^{-2}$\\
    monodisperse dust &&&\\\hline
    Aggregates composed of  & 10.3$\times10^{-5}$ & 2.1$\times10^{-5}$&2.0$\times10^{-2}$ \\
    polydisperse dust&&&\\\hline
    \end{tabular}
  \label{t:gamma}
\end{table}

The effective surface energy $\gamma_{\textrm{eff}}$ was measured, to be 8.0$\times10^{-5}$ and 10.3$\times10^{-5}$~J~m$^{-2}$ for aggregates composed of mono- and polydisperse dust, respectively (see Figure~\ref{f:gamma} and Table~\ref{t:gamma}). For these values, the surface energy of a monomer particle was calculated to be $\gamma$ = 1.7$\times10^{-2}$~J~m$^{-2}$ for the monodisperse and 2.0$\times10^{-2}$~J~m$^{-2}$ for the polydisperse SiO$_2$ particles. These values are in very good agreement with 2.5$\times10^{-2}$~J~m$^{-2}$ measured for silica powder by \citet{kendall_et_al1987Nature} and 1.86$\times10^{-2}$~J~m$^{-2}$ measured for $\sim$1 $\mu$m SiO$_2$ particles by \citet{heim_et_al1999PRL}. This confirms the validity of the scaling model developed in \citet{weidling_et_al2012Icarus}. This possible scaling from particles to aggregates could, for instance, be used in molecular dynamics simulations to investigate clusters composed of 100~$\mu$m-sized aggregates. Furthermore, it seems reasonable to try to adapt a collision recipe developed for clusters composed of $\mu$m-sized particles \citep{dominik_and_tielens1997ApJ} to clusters composed of $\sim$100 $\mu$m-sized aggregates.

\subsection{Application to protoplanetary disks}
\label{s:ppd}

Three commonly used protoplanetary disk models were derived by \citet{weidenschilling1977ASS}, \citet{andrews_and_williams2007ApJ}, and \citet{desch2007ApJ}, respectively. These models are based on surface density ($\Sigma$) and temperature $(T$) profiles of the form
\begin{equation}
\Sigma(\hat{r}) = \Sigma_0(\frac{\hat{r}}{1\textrm{ AU}})^{-\delta}
\end{equation}
and
\begin{equation}
T(\hat{r}) = T_0(\frac{\hat{r}}{1\textrm{ AU}})^{-\epsilon} ,
\end{equation}
where $\hat{r}$ is the distance to the central star, $\Sigma_0$ and $T_0$ are the values of the surface density and temperature at $\hat{r}=$1~AU, and $\delta$ and $\epsilon$ are the respective exponents of the power laws. All three models assume $T_0=280$~K and $\epsilon = 0.5$. Table~\ref{t:models} lists the surface density parametres. The Minimum Mass Solar Nebula (MMSN) model developed by \citet{weidenschilling1977ASS} is based on the minimum mass required in an original protoplanetary disk to build the Solar System as it exists today ($\Sigma_0=1700$~g/cm$^2$). \citet{andrews_and_williams2007ApJ} derived a low density nebula model from the astronomical observations of circumstellar dust disks in Taurus-Auriga ($\Sigma_0$ = 20~g/cm$^2$) and \citet{desch2007ApJ} a high density model by considering planet migration following their formation in a compact configuration around the newly formed star ($\Sigma_0$ = 50500~g/cm$^2$).

\begin{table}[t]
\caption{Model parametres for surface density of three protoplanetary disk models.}
\begin{center}
\begin{tabular}{|l|c|c|}
\hline
\textbf{PPD model} & {\boldmath $\Sigma_0$} \textbf{[g/cm$^2$]} & {\boldmath $\delta$} \\ \hline
\citet{weidenschilling1977ASS} & 1700                           &  1.5                             \\ \hline
\citet{andrews_and_williams2007ApJ}   & 20                           & 0.8                              \\ \hline
\citet{desch2007ApJ}   & 50500                          & 2.168                              \\ \hline
\end{tabular}
\end{center}
\label{t:models}
\end{table}

Turbulence in the disk is commonly assumed to follow a Kolmogorov cascade and the so-called $\alpha$-prescription \citep{shakura_and_sunyaev1973AA}, where the dimensionless parametre $\alpha = \tilde{\nu}/c_{\textrm{s}}H$ describes the strength of the turbulence, $c_{\textrm{s}}$ being the sound speed in the disk, and $\tilde{\nu}$ and $H$ the disk viscosity and height, respectively. Typical values for $\alpha$ are 10$^{-5}$ in quiet regions of the disk, e.g. in a "dead-zone" that could develop around 1~AU \citep{turner_et_al2007ApJ,brauer_et_al2008AAb}, and up to 10$^{-2}$ in more turbulent regions.

To compute the dust aggregate relative velocities for different aggregate sizes in the PPD midplane, an adaptation of a code developed by \citet{brauer_et_al2008AAa} was used. The effects considered to induce relative velocities were Brownian motion \citep{einstein1905AnP}, radial drift \citep{whipple1972PlP,weidenschilling1977MNRAS} and turbulence \citep{ormel_and_cuzzi2007AA}.

\begin{figure}[tp]
  \begin{center}
  \includegraphics[width = 0.40\textwidth]{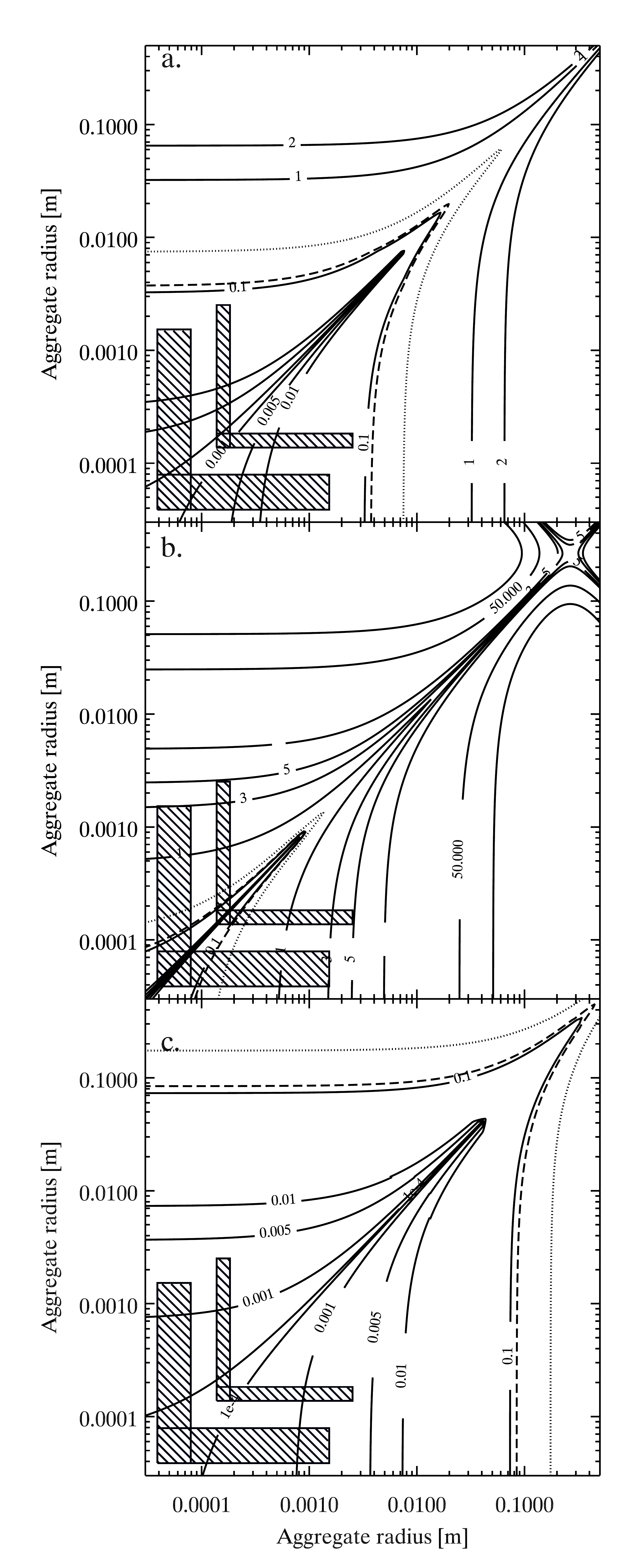}
 \caption{Relative velocities between dust aggregates in a protoplanetary disk computed according to a. \citet{weidenschilling1977ASS}, b. \citet{andrews_and_williams2007ApJ}, and c. \citet{desch2007ApJ} at 1~AU and a turbulence parametre of $\alpha = 10^{-5}$. The velocity profiles are labelled in units of m~s$^{-1}$. The hashed boxes represent the collisions observed during the SPACE experiment (see text for more details). The dashed contour displays the minimum sticking velocity measured during the SPACE experiment for both size distributions. The dotted contour displays the same minimum sticking velocity in the case of aggregate-aggregate collisions instead of aggregate-cluster collisions (see text for details).}
 \label{f:ppd_stick}
 \end{center}
\end{figure}

The computed relative velocity profiles for a low turbulence region ($\alpha$ = 10$^{-5}$) can be seen in Figure~\ref{f:ppd_stick} for all three models described above at 1~AU. The aggregate and cluster size ranges investigated in the SPACE experiment are represented by hashed boxes, to indicate which "natural" relative velocities the involved aggregates would possess in a protoplanetary disk. The heights of the horizontal boxes represent the initial size distributions of aggregates inserted into the experiment cells and the widths of the horizontal boxes span from the monomer aggregate size to the mean size of clusters during slow shaking phases (see Section~\ref{s:compare} for more details). In order to compare the relative velocities of aggregates in PPDs to the ones induced in the SPACE experiment, the minimum velocity for perfect aggregate sticking determined for both size distributions (see values in Table~\ref{t:vc}, dashed contour) was added to the figure.

As the collisions observed in the SPACE experiment occurred between individual aggregates and stiff aggregate clusters of infinite mass (see Section~\ref{s:cluster_morphology}), the reduced mass of these collisions is the same than the mass of an individual aggregate. The change in momentum during the collision is therefore $\Delta p = mv$. This value of $\Delta p$ could also be obtained by two aggregates of the same mass $m$ (implying $\mu=\frac{1}{2}m$) flying towards each other with the same velocity $v$ \citep[the relative velocity then would be $2v$, see][for more details]{brisset2014_diss}. In Figure~\ref{f:ppd_stick}, we also show the minimum sticking velocity between two same-size aggregates computed from our measurement of the sticking velocity between aggregates and stiff clusters (dotted line). 

We can now see that for the MMSN \citep[][Figure~\ref{f:ppd_stick}a.]{weidenschilling1977ASS} and the compact \citep[][Figure \ref{f:ppd_stick}c.]{desch2007ApJ} nebula models, the expected relative velocities for aggregate-aggregate and aggregate-cluster collisions are lower than the measured velocity for perfect aggregate sticking for the aggregates (both dashed and dotted lines). This indicates that at 1 AU and in a low-turbulence environment the investigated collisions would always lead to cluster growth. In the MMSN model, stiff clusters would stop growing through collisions with sub- to mm-sized particles when they reach about 4-mm in size (dotted line). The growth through same-sized aggregate-aggregate collisions, however, continues up to sizes of around 5~cm (dotted line in Figure~\ref{f:ppd_stick}a). For the compact nebula model \citep[][Figure \ref{f:ppd_stick}c.]{desch2007ApJ}, growth of stiff clusters through aggregate sweep-up would only lead to cluster sizes of up to 10~cm, while aggregate-aggregate collisions could grow clusters up to about 50~cm in size. In the low density model \citep[][Figure~\ref{f:ppd_stick}b.]{andrews_and_williams2007ApJ}, growth of stiff clusters through aggregate sweep-up would only lead to cluster sizes between 0.1 and 0.2~mm, while aggregate-aggregate collisions could grow clusters up to about 1~mm in size. Above these maximum size listed, bouncing is expected.

\section{Conclusion}

In this paper we presented the results of a multi-particle collision experiment conducted under reduced gravity conditions. The particles investigated were sub-mm-sized aggregates composed of micrometre-sized spherical and irregular SiO$_2$ dust particles. The experiment run on the REXUS 12 suborbital rocket provided 150 consecutive seconds of data observing the collision behaviour of these aggregates. The reduced gravity environment of the suborbital rocket was necessary to reach relative velocities between aggregates as low as <1~cm~s$^{-1}$ and to investigate the aggregate collision behaviour over a continuous range of relative velocities. During the SPACE experiment run, clusters of up to a few millimetres in size were formed. Owing to the rather high acceleration environment on the suborbital flight ($\gtrsim$1~m~s$^{-2}$, owing to the shaking of the experiment cell) the formed clusters restructured into monolayers through rolling of the aggregates upon sticking. Data analysis showed that this behaviour did not affect the measurements of the aggregate sticking probability and their pull-off force. 

A direct measurement of the sticking probability for impact velocities from about 3 to 22 cm~s$^{-1}$ was possible (Figure~\ref{f:beta}). The sticking probability of the dust aggregates was shown to be velocity dependant and the transition from a bouncing regime (sticking probability close to~0) to a sticking regime (sticking probability at~1) was very sharp.This indicates the existence of a sticking threshold for relative velocities between aggregates.

The suborbital flight also allowed for the study of the properties of sub-mm-sized aggregates built into clusters. The sequenced growth and fragmentation phases made it possible to measure the tensile strength and the effective surface energy of the aggregate clusters. The tensile strength values measured during the SPACE experiment are on the order of a few Pa, in agreement with independent measurements \citep{blum_et_al2014Icarus} and model predictions \citep{skorov_and_blum2012Icarus}. The value for the effective surface energy of the aggregates is in very good agreement with the surface energy measured for $\sim$1~$\mu$m particles \citep{heim_et_al1999PRL}, combined with the scaling model of \citet{weidling_et_al2012Icarus}. This indicates that dust aggregates can be used as monomer particles in molecular dynamics simulations, with a proper scaling of the particle parametres. This allows for simulations of clusters up to 10~cm in size with the computational capacities currently used.

As shown in Figure~\ref{f:ppd_stick}, the measured sticking threshold for the aggregates studied in this work leads to different dust growth behaviour predictions depending on the protoplanetary disk model considered. For a disk dust surface density at 1~AU of 1700~g~cm$^{-2}$ \citep{weidenschilling1977ASS}, the results of hte experiment presented in this work indicate a dust grain growth to sizes of about 1~cm at 1~AU. In the case of a lower density model \citep{andrews_and_williams2007ApJ}, dust grains would grow only to a few 100~$\mu$m, while a higher density model \citep{desch2007ApJ} would allow for uninterrupted dust grain growth. 

\section*{Acknowledgements}
We thank the REXUS/BEXUS project of the Deutsches Zentrum f{\"{u}}r Luft- und Raumfahrt (DLR) for the flight on the REXUS 12 rocket. This work was supported by the Interaction in Cosmic and Atmospheric Particle Systems (ICAPS) project of DLR (grants 50WM0936 and 50WM1236) and a fellowship from the International Max Planck Research School on Physical Processes in the Solar System and Beyond (IMPRS) at the Universities of Braunschweig and G\"ottingen. We also thank Dipl. Phys. Oliver Lenck from the Frauenhofer Institute for Surface Engineering and Thin Films of Braunschweig for the anti-adhesive glass coating of the particle containers.

\appendix

\section{Analysis methods}
\label{a:analysis}

\paragraph{Frame correction}

\begin{figure}[t]
  \begin{center}
  \includegraphics[width = 0.48\textwidth]{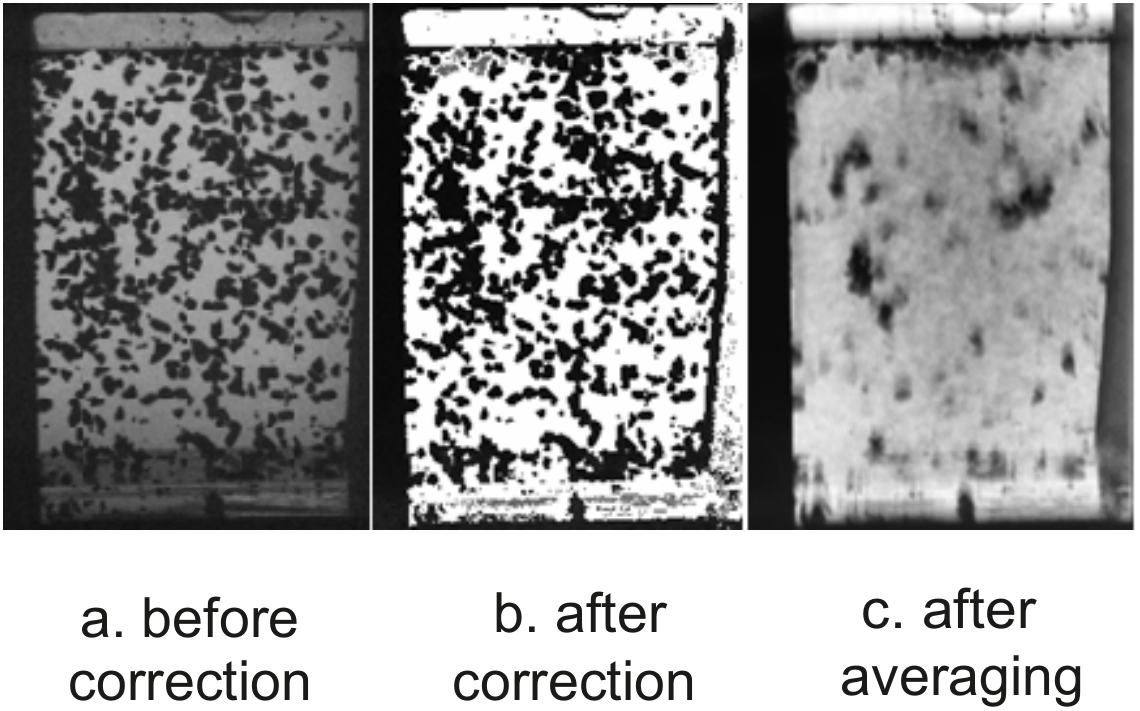}
 \caption{Image processing steps. a. Recorded SPACE data frame at 11.8~s after beginning of frame recording, showing one of the smaller experiment cells. b. The same frame after correction of background illumination. c. The same frame after background correction and averaging over 201 frames. The clusters on the cell walls become apparent.}
 \label{f:processing}
 \end{center}
\end{figure}

\begin{figure}[tp]
  \begin{center}
  \includegraphics[width =0.45\textwidth]{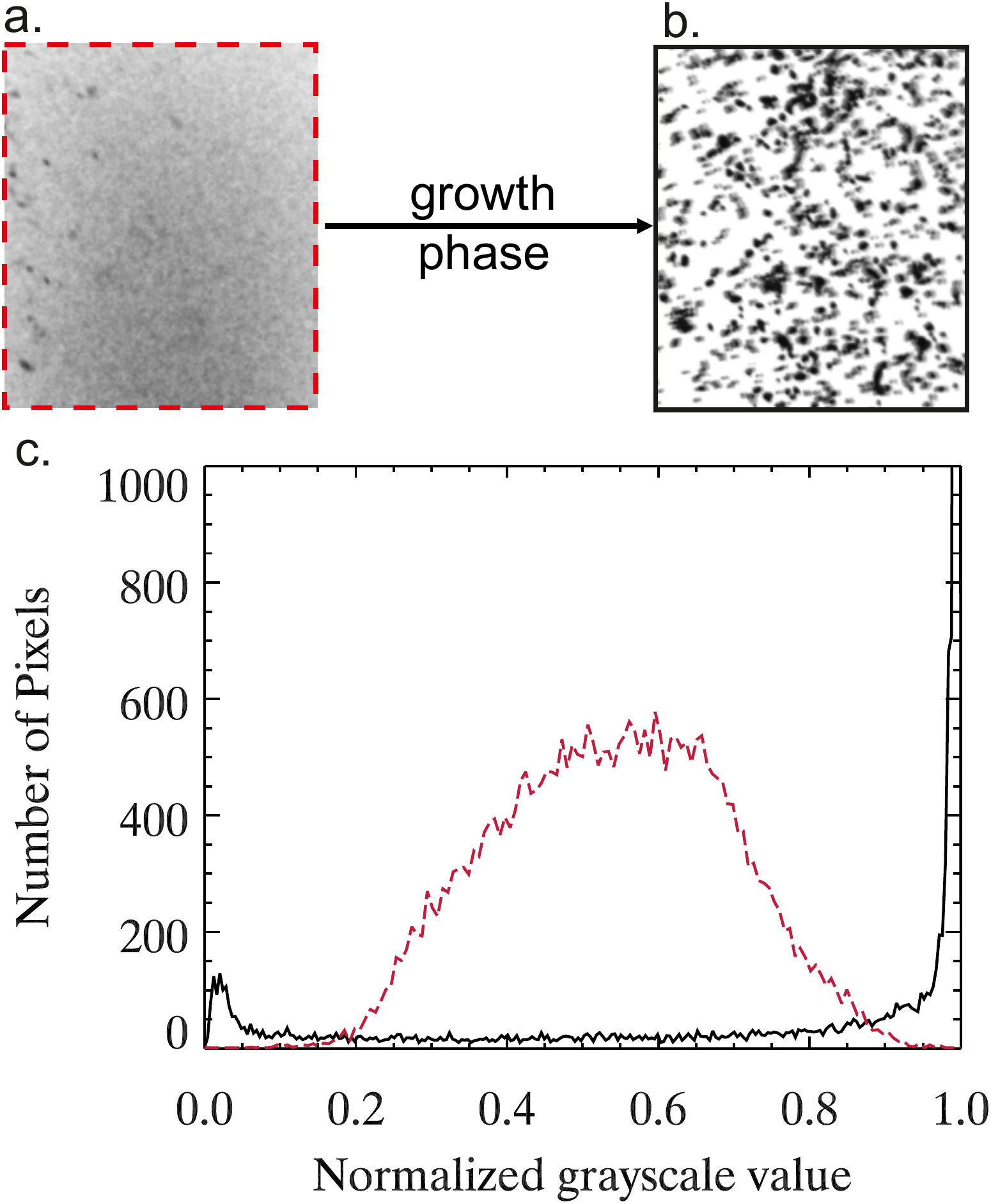}
 \caption{Analysis method used on the SPACE experiment data to determine the number of free-flying aggregates in the experiment cells at each moment in time. a. Region of interest inside an averaged data frame of the larger experiment cell (see Figure~\ref{f:cells}b.) at the beginning of the experiment run (1.2~s after start of recording), when all aggregates were free-flying. Image processing as shown in Figure 7c was applied. b. Region of interest inside an averaged data frame of the same experiment cell during the slowest shaking phase (47.1~s after start of recording), when all aggregates are incorporated in clusters sticking to the glass walls. Image processing as shown in Figure 7c was applied. c. greyscale histograms of frames a. and b. When all aggregates are free-flying (a.), the background is uniformly grey (dashed red line), whereas the dark clusters (small peak at lower greyscale values) and the bright background (high and narrow peak at higher greyscale values) can clearly be distinguished when the aggregates are all incorporated in clusters (solid black line). }
 \label{f:analysis_gro}
 \end{center}
\end{figure}

At the beginning of the data analysis sequence, the recorded frames were corrected as follows:
\begin{itemize}
\item Elimination of the rotational movement of the experiment cells. This was done by detecting the cell limits in each frame and shifting the frame so that these limits were always in the same position. In these corrected frames, the free-floating particles possess cycloid motions instead of linear trajectories. The cells had an additional movement along the line of sight of the camera because they were slightly wobbling due to friction between the cog wheels of the shaking mechanism. This produced a non-zero noise level in the greyscale histogram of averaged frames (see Figure \ref{f:analysis_gro}c.) that could be disregarded as not being related to the number of free-flying aggregates.
\item Correction of spatial variations in background illumination. The background illumination was not perfectly even over the field of view of the camera. To rectify this, the average background was determined over all the recorded frames and the mean value then subtracted from each frame (see Figure~\ref{f:processing}b.).
\item Checking for temporal variations in the transmission of the glass windows. As the number of collisions between the free-flying dust aggregates and the glass walls was very high and such collisions could pollute the glass walls with small dust particles, it was necessary to check for the glass cleanliness over the experiment run. This was done by picking a few random positions on a frame and checking their greyscale values at several moments in time during the experimental run when they were not covered by an aggregate. The result of this investigation was that the glass stayed clean in each of the three experiment cells confirming the low glass sticking efficiency determined in Appendix~\ref{a:mc}. This also means that the anti-adhesive coating worked well on more than 99\% of the glass surfaces and that the deposition of seed aggregates on the glass walls was probably due to small failures of the anti-adhesive coating.
\end{itemize}
The thus corrected frames were then used for the analysis of aggregate collisions and growth.

\paragraph{Frame averaging}
\label{s:averaging}
Because of the high optical depth of the free-flying aggregates in the experiment cells, it was not possible to measure their number and individual sizes directly. A way of still accessing this information was to perform a running average of the image frames. Because the free-floating aggregates were constantly moving, the total cross-sectional area of each aggregate was thus taken into account by contributing to the background greyscale. Furthermore, non-moving clusters growing on the glass walls became fully visible. Thus, each frame of the SPACE experiment was averaged over 201 frames, i.e. over a time of 1.2~s.
An example of the frames obtained by this averaging can be seen in Figure~\ref{f:processing}c. The clusters on the glass walls are now apparent as dark spots and can be monitored. Furthermore, the background greyscale value of cluster-free parts of the frames delivers information on the number of dust aggregates that are free-flying, i.e. the darker the background, the more aggregates are free-flying, and the brighter the background, the more aggregates are incorporated in clusters on the cell walls.

\paragraph{Statistical analysis methods}
The analysis of the frame background greyscale and of the clusters on the cell walls are in fact two distinct ways of retrieving information on the number and mass of the aggregates and clusters in the experiment. Figure \ref{f:analysis_gro} shows histograms of two averaged data frames. In the left frame (a.), all of the dust aggregates present in the cell are free-flying. Its greyscale histogram (c., dashed red line) shows a rather diffuse distribution of medium-dark greyscale values. In the right frame (b.), however, all aggregates are incorporated in clusters and none are free-flying. The histogram now shows two distinct peaks (c., solid black line), a strong and narrow peak for high greyscale values accounting for the now very bright (corrected) background, and a smaller but also quite narrow peak at very low greyscale values accounting for the dark visible clusters on the cell walls. As the glass remained perfectly clean during the experiment run, these two frames can be used to calibrate the background greyscale to the fraction of aggregates free-flying in the cell volume at this time (see Section~\ref{s:n}).

\begin{figure}[tp]
  \begin{center}
  \includegraphics[width = 0.45\textwidth]{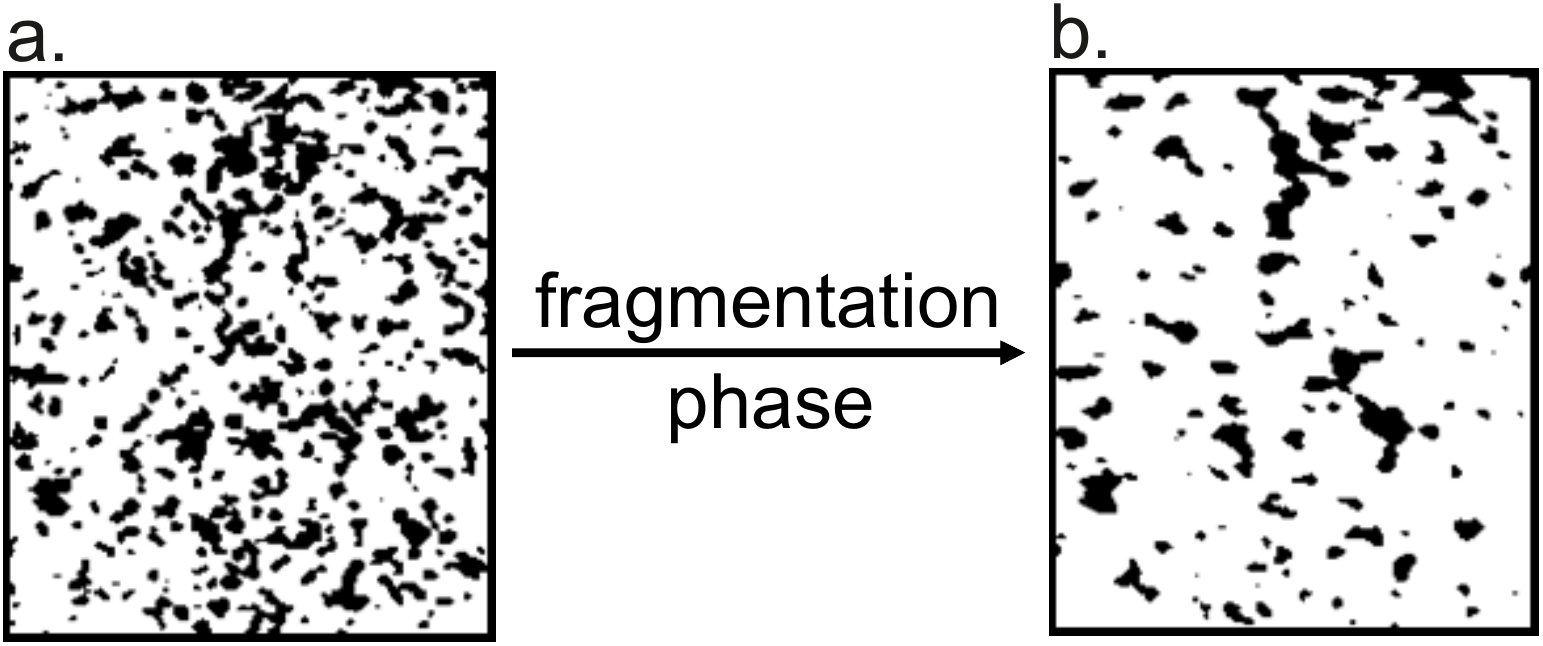}
 \caption{Analysis of the clusters on the cell walls inside the SPACE experiment during the suborbital flight. a. Region of interest inside an averaged and binarized data frame of the bigger experiment cell at the beginning of a fragmentation phase (61.8~s after beginning of the recording). b. Region of interest inside an averaged and binarized data frame of the same cell at the end of this fragmentation phase (76.5~s after beginning of the recording). The area covered by clusters (black areas) has visibly decreased.}
 \label{f:analysis_frag}
 \end{center}
\end{figure}

Another analysis method is to binarize the averaged frames and monitor the area covered by clusters at a certain point in time. Figure \ref{f:analysis_frag} illustrates this by presenting two averaged and binarized frames at the beginning and the end of a fragmentation phase. Clusters (black areas) have partly dissolved over time. The total surface covered by clusters can be tracked during the experiment run (see Section~\ref{s:area}).

\section{Investigation of the cluster growth by aggregate-aggregate collisions}
\label{a:mc}

This appendix describes the Monte Carlo simulations performed depositing particles on a surface representing the glass walls of the SPACE experiment cell. The purpose of these simulations was to reproduce the aggregate deposition pattern seen on the cell walls and to investigate whether all aggregates stuck directly to the glass or tended to form clusters through dust-dust collisions. The parametres that were varied between the simulations were the initial number of particles to be deposited on the glass surface, $N_{\textrm{sim}}$, and a sticking probability for aggregate-glass collisions, $\kappa$.

Before the simulations are started, no particles are deposited on the surface. A number $N_{\textrm{sim}}$ of particles are then created with a size distribution similar to the size distribution of the dust aggregates and inserted into the experiment cells. At each step of the simulation, one of the $N_{\textrm{sim}}$ particles collides with a randomly chosen location on the surface. A collision with the bare glass surface either leads to a deposition of the aggregate according to the sticking probability $\kappa$, or to bouncing back into the cell volume. In the latter case, the particle collides with the surface again, on a new random position. A collision with another particle already sticking to the surface always leads to deposition (a sticking probability of unity is considered for aggregates impinging those sticking to the glass wall). The simulation stops when all $N_{\textrm{sim}}$ particles are deposited.

As the frames recorded during the experiment had a resolution of 56.5~$\mu$m/pixel, it was possible to simulate the particle growth on a one-to-one scale. The smaller sieved aggregates observed in the SPACE experiment had a mean size of 120~$\mu$m ranging from $\sim$20~to~300~$\mu$m in diametre and the bigger aggregates had a mean size of 330~$\mu$m ranging from $\sim$50~to~600~$\mu$m (see Figure~\ref{f:dust_distrib}). Hence, 1 to 81 pixel particles inserted into the simulation correspond to the size range of dust aggregates observed in the SPACE experiments. Simulating the wall surfaces by 221$\times$362~pixel and 192$\times$141~pixel corresponds to regions of interest inside the 15$\times$24~mm$^2$ and 15$\times$11~mm$^2$ visible surfaces of the larger and smaller experiment cells.

For the small monodisperse SiO$_2$ dust distribution, the real inserted number of aggregates was $N_{\textrm{exp}}\simeq4000$. Thus, the initial number of particles created in the simulations was varied from $N_{\textrm{sim}}=3000$ to $N_{\textrm{sim}}=5000$, with a step size of 100 particles. For the larger size distributions of aggregates composed of mono-and polydisperse dust, the real number of inserted aggregates was $N_{\textrm{exp}}\simeq400$. Here, the initial number of particles was varied from $N_{\textrm{sim}}=300$ to $N_{\textrm{sim}}=500$ with a step size of 10 particles. For both distributions, the glass sticking parametre $\kappa$ was varied from $\kappa=0.0001$ to $\kappa=0.1$ with a linear step size of 0.0001.

\begin{figure*}[p]
  \begin{center}
  \includegraphics[width = 0.85\textwidth]{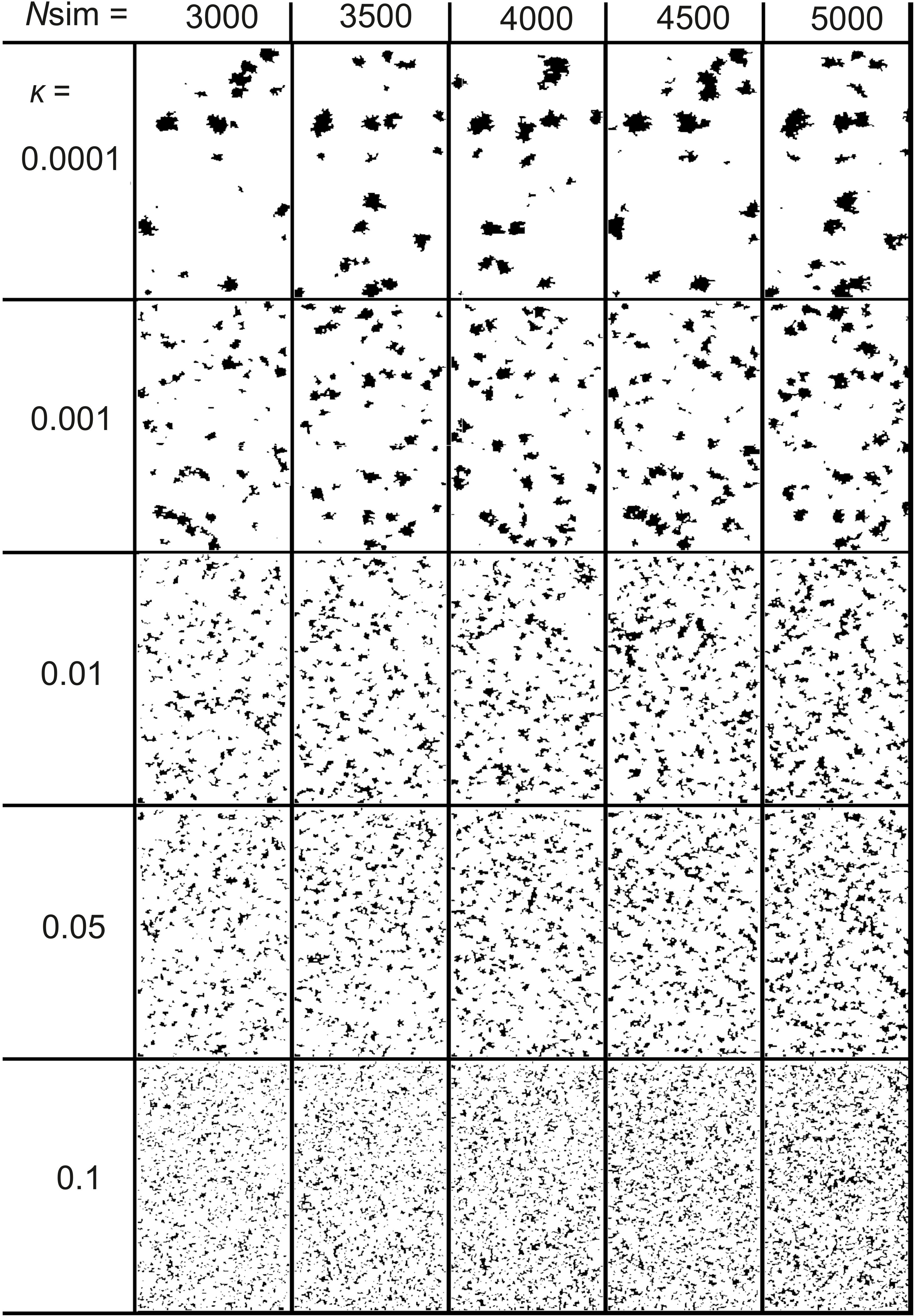}
 \caption{Summary of the Monte Carlo simulation results for the agglomeration of the small monodisperse dust size distribution used in the SPACE experiment rocket flight (mean size of 120 $\mu$m). The simulation surface of 221$\times$362~pixel represents a region of interest inside the bigger experiment cell. Each column has the same initial number of particles inserted into the simulation $N_{\textrm{sim}}$. Each row has the same glass sticking probability $\kappa$.}
 \label{f:mc_small}
 \end{center}
\end{figure*}

\begin{figure*}[tp]
  \begin{center}
  \includegraphics[width = 0.98\textwidth]{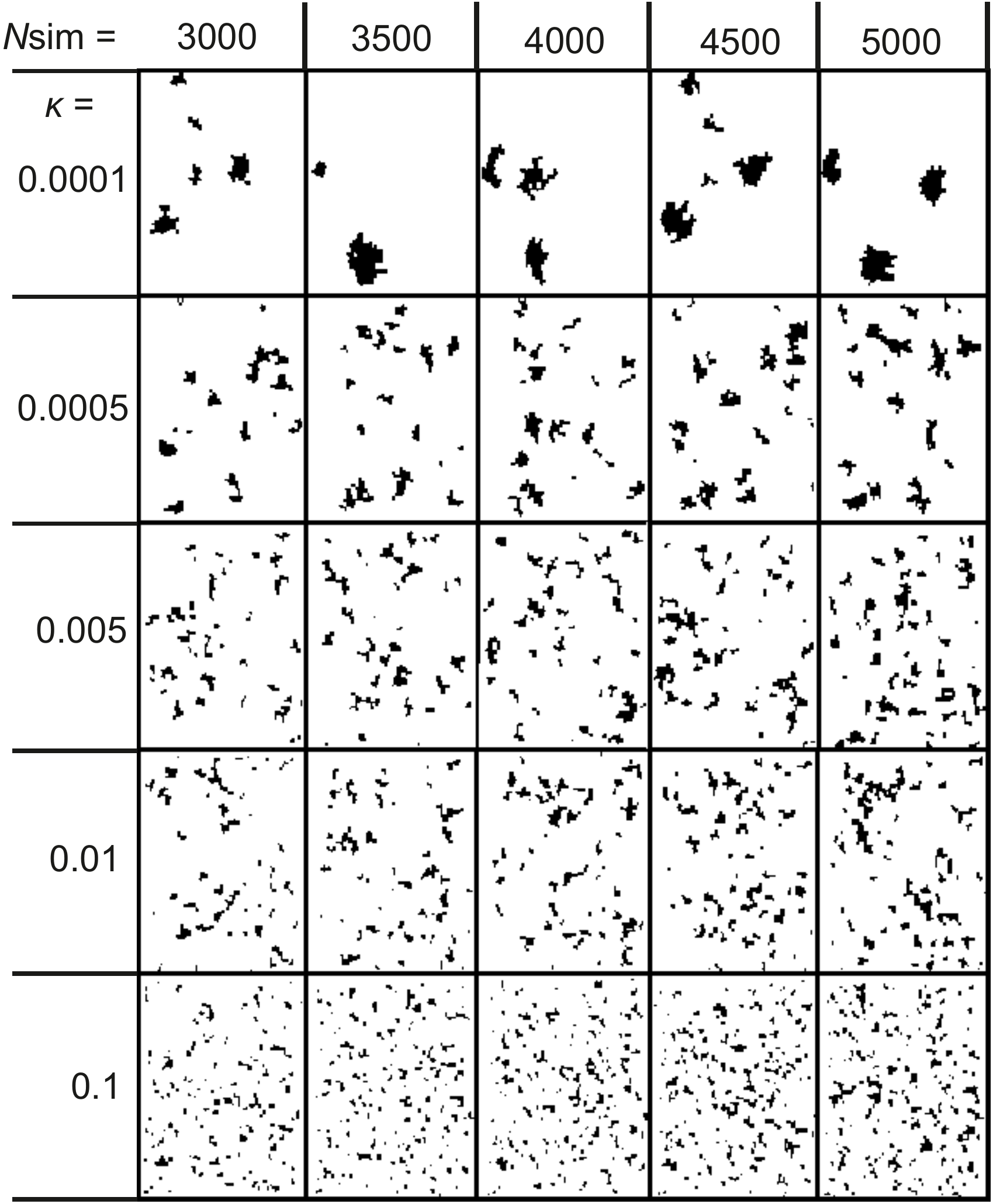}
 \caption{Monte Carlo simulation results for the agglomeration of the large size distribution of aggregates composed of monodisperse dust used in the SPACE experiment rocket flight (mean size of 330 $\mu$m). The simulation surface of 192$\times$141~pixel represents a region of interest inside one of the smaller experiment cells. Each column has the same initial number of particles inserted into the simulation $N_{\textrm{sim}}$. Each row has the same glass sticking probability $\kappa$.}
 \label{f:mc_big}
 \end{center}
\end{figure*}

A few examples of the simulation results can be seen in Figures~\ref{f:mc_small} and~\ref{f:mc_big}. For low sticking probabilities $\kappa$, only very few seeds deposited on the surface and clusters grow from there, leading to a few large clusters. For higher values of $\kappa$, many more particles deposit on the glass surface and the simulation results in a higher number of smaller clusters.
To quantitatively compare the results of the simulation with the experimental data, a data frame taken at 47~s after start of recording was used. This frame was recorded during a slow shaking phase (see Figure~\ref{f:motor_profile}), while all particles were incorporated in clusters and none were free-floating. This frame was binarized and the resulting number of clusters and their size distribution were determined. The root mean square of the bin-by-bin difference between the measured cluster size distribution and the size distribution of the simulation result, $\sigma_{\textrm{dist}}$, was used as the comparison criterium. To smooth out binarization effects on the size distribution of the data frame, the bin sizes were chosen at 4~pixel for the small and 9~pixel for the larger dust distribution. The simulations were run 5000~times for each set of parametres and the mean $\sigma_{\textrm{dist}}$ determined.

\begin{figure}[tp]
  \begin{center}
  \includegraphics[width = 0.48\textwidth]{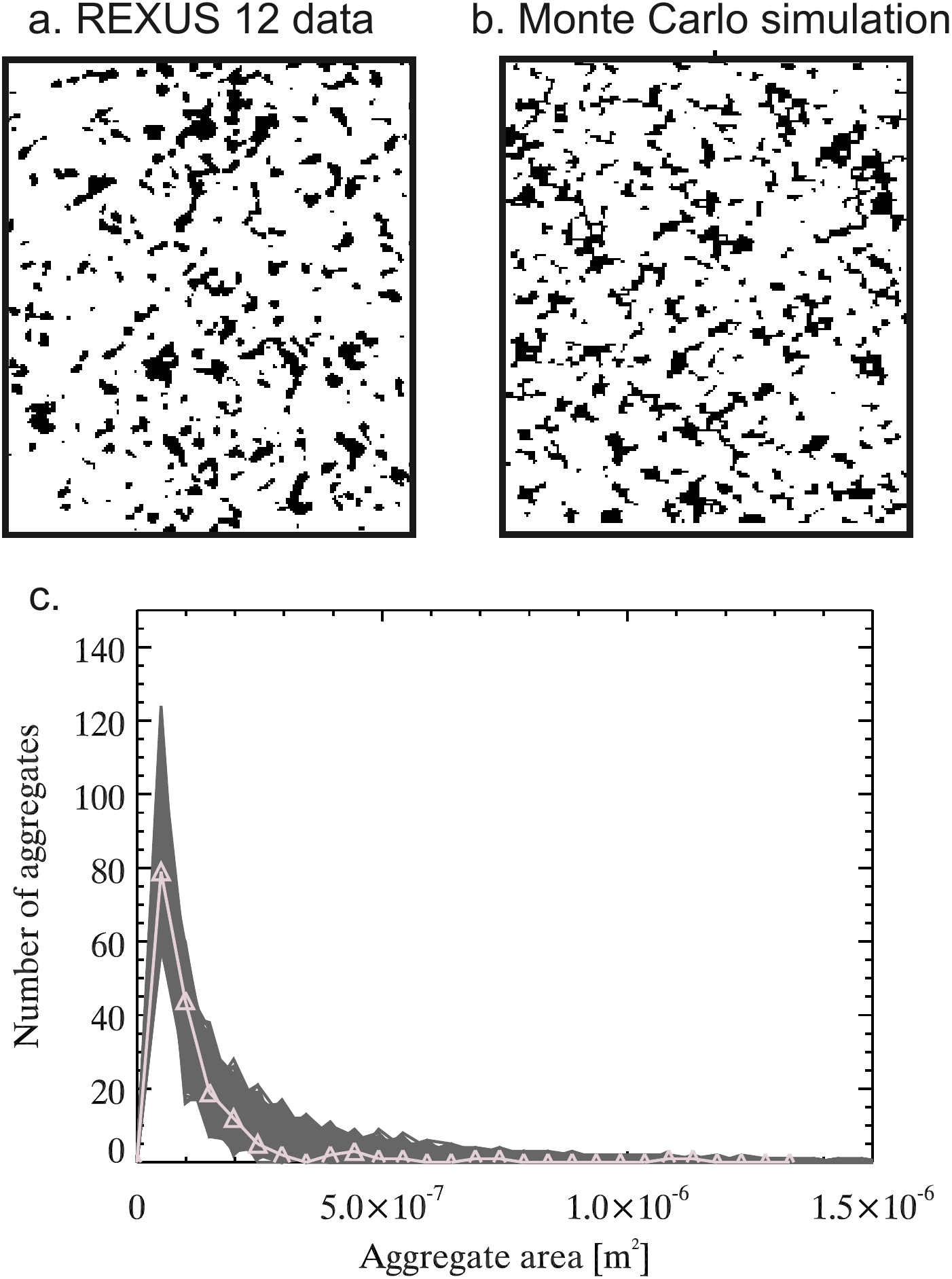}
 \caption{Results of a Monte Carlo simulation for the smaller aggregate size distribution used in SPACE during the suborbital flight: a. binarized SPACE data frame at 47 s after start of data recording. At this point of the experiment run, all dust aggregates are clustered on the cell wall. b. Monte Carlo simulation depositing 4600 particles on a surface with a glass sticking parametre $\kappa$ = 0.0055, c. size distribution of the data frame seen in a. (light red line, triangles) and 5000 runs of Monte Carlo simulations with the parametres of image b. (grey lines). The bin size for these distributions is 4 pixel (1.23$\times10^{-8}$m$^2$).}
 \label{f:mc}
 \end{center}
\end{figure}

The best matches (minimum $\sigma_{\textrm{dist}}$) were found for $N_{\textrm{sim}}=4600$ and $N_{\textrm{sim}}=360$ and $\kappa=0.0055$ and $\kappa=0.0045$ for the smaller and larger aggregate distributions, respectively (Table~\ref{t:mc}). Figure~\ref{f:mc} illustrates the best match between data and simulations for the smaller aggregate size distribution. The size distributions of 5000~simulation runs and of the measured data frame are plotted in grey and in light red, respectively. It can be seen that the size distribution of the data frame lies within the size distribution range obtained for the simulations. As the number $N_{\textrm{sim}}$ of inserted particles is in good agreement with the number of aggregates inserted into the SPACE experiment cells, glass sticking values around 0.5~\% can be assumed.

It can be concluded that it was not an enhanced glass sticking efficiency that led to clusters on the cell walls but indeed the enhanced number of collisions of the aggregates on the glass, induced by residual accelerations during the rocket flight (see \ref{a:res_acc}). This is also supported by the fact that the glass walls did not get dirty during the experiment run despite the very high number of collisions. It can also be concluded that dust clusters were not growing on the glass directly but on a few dust aggregates that had stuck to failure points of the anti-cohesive coating of the glass walls and then served as seeds.
\begin{table}[t]
\caption{Monte Carlo simulation results: the best match between simulation and data for both dust distributions are presented. $N_{\textrm{exp}}$ and $N_{\textrm{sim}}$ are the numbers of aggregates inserted into the experiment cell and into the simulation, respectively, $\kappa$ is the glass sticking probability parametre.}
\begin{center}
\begin{tabular}{|l|r|r|r|}
\hline
\textbf{Aggregate distribution} & \textbf{$N_{\textrm{exp}}$} & \textbf{$N_{\textrm{sim}}$} & \textbf{$\kappa$}\\ \hline
Small & 4240 & 4600 & 0.0055\\ \hline
Large & 375 & 360 & 0.0045\\ \hline
\end{tabular}
\end{center}
\label{t:mc}
\end{table}


\section{Calculating perturbation accelerations during the suborbital flight}
\label{a:other_influences}

\subsection{Residual accelerations during the rocket flight}
\label{a:res_acc}
During the suborbital flight of SPACE, the dust aggregates in the containers were subjected to two types of residual acceleration, caused by atmospheric drag and rocket spin.

The residual atmospheric drag accelerations were due to the fact that the REXUS~12 rocket had an apogee of 82~km, an altitude at which the aerodynamic effects of the remaining atmosphere on the rocket are still influential. The drag force induced along the direction of flight is $F_{\textrm{drag}}=\frac{1}{2}\rho v^2AC_{\textrm{d}}$, with $\rho$ being the local air pressure, $v$, $A$ and $C_{\textrm{d}}$ the rocket's flight velocity, cross section and drag coefficient, respectively. Comparing the rocket's flight altitude profile with the recorded SPACE data, it could be determined that the streaming of the image frames started at an altitude of about 70~km, lasted through the passing of apogee (82~km) and stopped during descent at about 50~km altitude. Hence, it can be assumed that the relevant data to used for this analysis were recorded above an altitude of 70~km (the last seconds of recording were not relevant for the data analysis). The standard atmosphere model (1976 US Standard Atmosphere) at this altitude yields $\rho_{\textrm{air}} = 7.42\times10^{-5}$~kg/m$^3$. The cylindrical rocket had a diametre of $d$~=~0.356~m and, hence, a cross section of $A=\pi(\frac{\textrm{d}}{2})^2=9.95\times10^{-2}$~m$^2$. The drag coefficient $C_{\textrm{d}}$~=~0.341 was calculated from measurements made on a REXUS flight in 2009 made by \citet{anderson_et_al2009}. The maximum rocket speed was $v$ = 562~m~s$^{-1}$ at 70~km altitude, which leads to a maximum drag force on the rocket of $F$ = 0.398~N. For a rocket mass of $m_{\textrm{r}}$~=~515~kg, the maximum residual drag acceleration was $a_{\textrm{drag}}$ = $F_{\textrm{drag}}/m_{\textrm{r}}$ = 7.72$\times$10$^{-4}$~m~s$^{-2}$ = 7.86$\times$10$^{-5}g$, with $g$ being the Earth's gravitational acceleration. As the dust aggregates were free-flying in vacuum inside of the experiment cells, which are directly coupled to the rocket itself, this is the maximum residual drag acceleration they experienced relative to the cell walls.

The residual angular velocity of the rocket after de-spin was measured by the Service Module to be $\omega_{\textrm{r}}$ =~0.19 rad/s. The resulting centrifugal acceleration on one of the SPACE dust aggregates was, hence, $a_{\textrm{centrif}}=\omega_{\textrm{r}}^2R$ with $R$ being the distance of the aggregate to the roll axis of the rocket. The way the experiment was designed, the distance to the rocket's roll axis could be approximated to $R$~=~0.04~m. Thus, the residual spin acceleration on the aggregates was about $a_{\textrm{centrif}}=1.47\times10^{-4}g$.

The effects of these residual accelerations on the behaviour of the dust aggregates can be seen in Figure~\ref{f:frames}b. The tendency of the aggregates to gather in the upper left corner of the frame reveals the combination of a residual acceleration acting on them in the direction of flight (vertical on the image frames) and in the radially outward direction (horizontal on the image frames), increasing the number of their collisions in this corner of the cell.

To investigate whether these two residual accelerations had an influence on the collision behaviour of the observed aggregates, the acceleration values were compared to the those induced by the rotating cell walls. The minimum rotation frequency of the cells was $f_{\textrm{min}}$~=~4.78 Hz, inducing a minimum wall acceleration of $a_{\textrm{min}}$ = (2$\pi f_{\textrm{min}})^2\tilde{r}$ =  0.90~m~s$^{-2}$ =  9.2$\times10^{-2}g$, with $\tilde{r}$~=~1~mm being the rotation radius of the experiment cells. This is almost 3 orders of magnitude stronger than the residual accelerations calculated above. Therefore, it is reasonable to neglect the effects of residual accelerations in the data results.

\subsection{Magnetic and electrostatic effects}

To validate our results as being meaningful for protoplanetary dust aggregation, it is also important to rule out magnetic or electrostatic effects on the collision properties. Both magnetic and electrostatic effects are known to play an important role on dust particle collisions \citep[e.g.][]{poppe_et_al2000aApJ,Nubold_et_al2003Icarus}. However, in this series of experiments with SiO$_2$ dust aggregates, the intention was to investigate aggregate properties free of these influences.

Ruling out magnetic effects is trivial as none of the materials used for this experiment were magnetic.

Furthermore, to avoid electrostatic charging of any kind, the glass cells holding the particles were built inside of a frame acting as a Faraday cage and preventing external electric fields penetrating inside the cells. Before inserting the dust aggregates into the containers, they were grounded for several days to avoid initial particle charging.

\begin{figure}[t]
  \begin{center}
  \includegraphics[width = 0.45\textwidth]{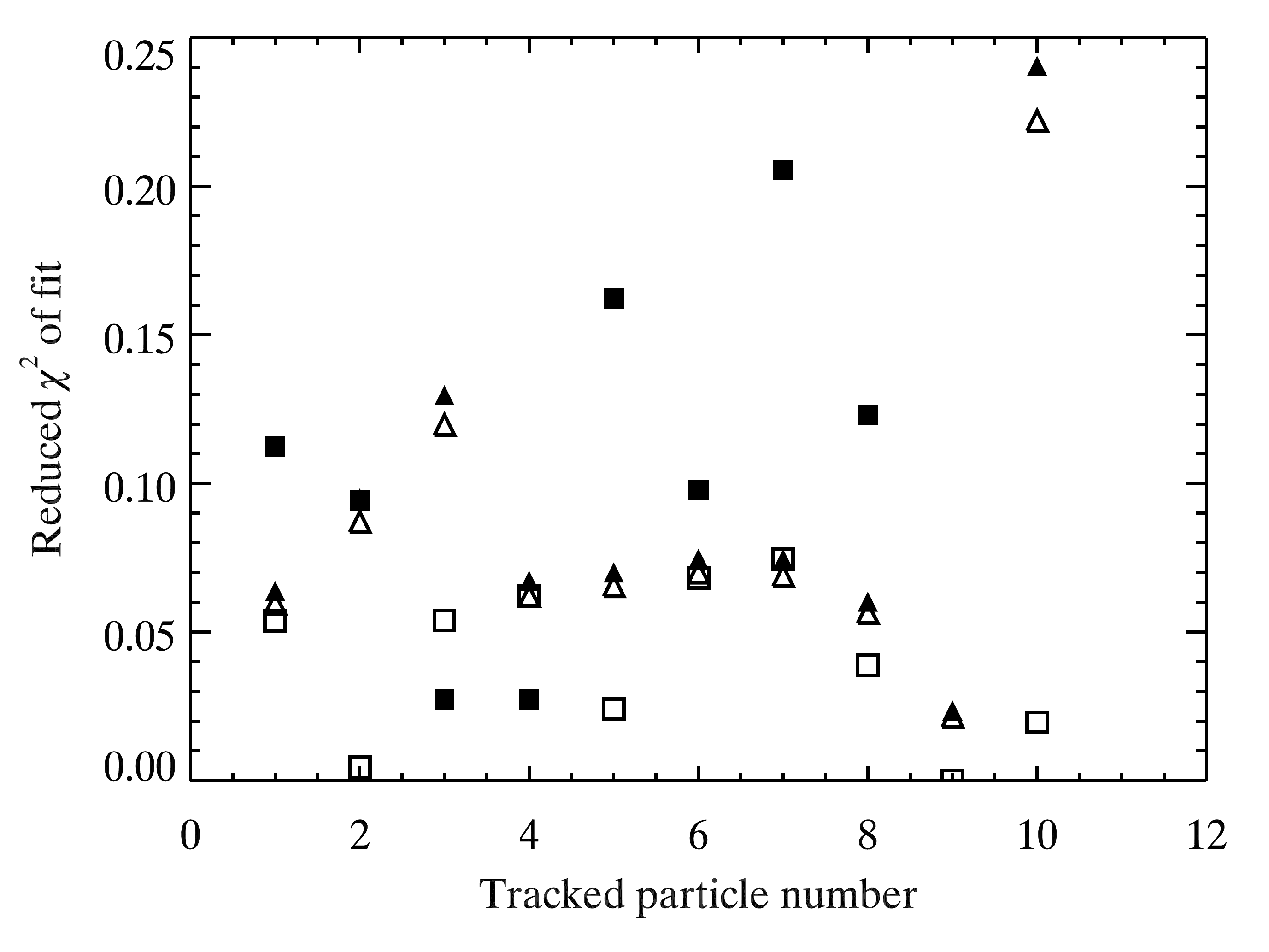}
 \caption{Reduced $\chi^2$ error statistics for linear and parabolic fits to the trajectories of 10 trackable aggregates in the cell containing the larger aggregates composed of monodisperse dust during the SPACE rocket flight experiment run, before (triangles, open for the linear fit and filled for the parabolic fit) and after impact with a glass wall (squares, open for the linear fit and filled for the parabolic fit).}
 \label{f:elec_check}
 \end{center}
\end{figure}

To make sure that these precautions guaranteed negligible electrostatic effects, the charge state of the aggregates observed during the SPACE flight was derived from the analysed data. During the SPACE experiment run, the relevant collisions were between free-flying aggregates and clusters forming on the cell walls. Electrostatic effects were ruled out by picking a few random aggregates, which could be directly observed colliding with a container wall, and determining the form of their trajectories. In the case of electrostatic charging of the dust and the cell walls, the aggregates would behave like charged particles in an electric field: their trajectories approaching the cell wall would be non-linear with respect to time due to induced acceleration or deceleration. For the 10 analysed collisions, the reduced $\chi^2$ error statistics of both the linear and parabolic fit of their trajectories approaching and leaving the glass wall was measured (see Figure~\ref{f:elec_check}). The mean $\chi^2$ error statistics for a linear trajectory fit were 0.083 and 0.040 before and after the aggregate collision with the cell wall, respectively. For the parabolic fit, these values were 0.90 and 0.12, respectively, indicating that there was no non-linearity in the particle trajectories.
The second degree coefficient of the formal parabolic fit, $p_2$, can be related to an electrical charge of the corresponding aggregate. Each aggregate is considered as being a charged particle of charge $q$ moving in a uniform electric field $E$, then $p_2 = \frac{1}{2}\frac{qE}{m}$, with $m$ being the mass of the aggregate. The electric field is approximated to $E$ = $\frac{q}{4\pi\epsilon_0r^2}$ with $\epsilon_0$ being the dielectric constant in vacuum and $r$ the distance from the aggregate to the interacting charge (taken here as $\sim$2~$\times$~aggregate radius). The mean number of elementary charges carried by each monomer particle ($\sim$1 $\mu$m in size) inside the aggregate is thus calculated to be 0.015. The induced accelerations are a$_{\textrm{el}}\, =\, 7.27\times10^{-11}$~m~s$^{-2}$. These accelerations can be neglected compared to the ones induced by the cell shaking.

\end{document}